\documentclass[aps,pra,showpacs,reprint,superscriptaddress]{revtex4-1}
\usepackage{amsmath}
\usepackage{amssymb}
\usepackage{mathtools}
\usepackage{graphicx}
\usepackage[colorlinks=true,urlcolor=blue,menucolor=blue,citecolor=black,linkcolor=blue]{hyperref}

\renewcommand{\vec}[1]{\mathbf{#1}}
\newcommand{\vecS}[1]{\boldsymbol{#1}}
\renewcommand{\L}{L}
\newcommand{\R}{R}
\newcommand{\C}{C}
\newcommand{\U}{U}
\newcommand{\MU}{\rm MU}
\newcommand{\BZR}{\mathcal{BZR}}
\newcommand{\BZ}{\mathcal{BZ}}
\newcommand{\br}[1]{\bar{#1}}
\newcommand{\tld}[1]{#1}

\newcommand{\inte}{\rm{eff}}

\newcommand{\AFI}{\rm{AFI}}
\newcommand{\AF}{\rm{AF}}
\newcommand{\NM}{\rm{NM}}
\newcommand{\Int}{\rm{Int}}

\newcommand{\TB}{\rm{TB}}
\newcommand{\Quad}{\rm{Q}}

\newcommand{\B}{\text{B}}
\newcommand{\Sh}{\text{Sh}}
\newcommand{\nearnb}[2]{\left< #1 , #2 \right>}
\DeclarePairedDelimiter\abs{\lvert}{\rvert}

\DeclareMathOperator{\sign}{sgn}

\begin{document}
\title{Superconductivity at metal-antiferromagnetic insulator interfaces}
\author{Eirik L\o{}haugen Fj\ae{}rbu}
\thanks{Currently at the Norwegian Defence Research Establishment (FFI), NO-2027 Kjeller, Norway; eirik-lohaugen.fjarbu@ffi.no}
\affiliation{Center for Quantum Spintronics, Department of Physics, Norwegian University of Science and Technology,
NO-7491 Trondheim }
\author{Niklas Rohling}
\affiliation{Center for Quantum Spintronics, Department of Physics, Norwegian University of Science and Technology,
NO-7491 Trondheim }
\affiliation{Department of Physics, University of Konstanz, 78457 Konstanz, Germany}
\author{Arne Brataas}
\affiliation{Center for Quantum Spintronics, Department of Physics, Norwegian University of Science and Technology,
NO-7491 Trondheim }

\begin{abstract}
Magnons in antiferromagnetic insulators couple strongly to conduction electrons in adjacent metals. We show that this interfacial tie can lead to superconductivity in a tri-layer consisting of a metal sandwiched between two antiferromagnetic insulators. The critical temperature is closely related to the magnon gap, which can be in the THz range. We estimate the critical temperature in Mn$\text{F}_2$-Au-Mn$\text{F}_2$ to be on the order of $1$\,K. The Umklapp scattering at metal-antiferromagnet interfaces leads to a d-wave superconductive pairing, in contrast to the p-wave superconductivity mediated by magnons in ferromagnets. 
\end{abstract}
\maketitle

\section{Introduction}

Antiferromagnetic insulators (AFIs) offer several advantages over ferromagnets such as higher operating frequencies and the absence of stray magnetic fields \cite{NatPhys.14.200, NatPhys.14.213}. 
Spin waves and their quanta, magnons, in AFIs couple strongly to electrons in adjacent normal metals (NMs) \cite{PhysRevLett.113.057601,PhysRevB.90.094408,PhysRevB.95.144408}. 
Importantly, this enables electric control of the antiferromagnetic spin dynamics. 
Even so, AFIs have received less attention than ferromagnetic insulators (FIs) in spintronics. 
A standard model for the interfacial tie is an exchange coupling between the itinerant electrons and the localized spins \cite{cheng2014aspects,PhysRevB.90.094408,PhysRevB.95.144408}. 
In this formalism, the electrons experience a staggered field and scatter through two different scattering channels: a regular channel and an Umklapp channel \cite{PhysRevB.90.094408,PhysRevB.95.144408}. 

In this paper, we show that the electron-magnon coupling at the NM-AFI interfaces can lead to superconductivity. 
The magnons in the AFIs mediate the superconductive pairing of the itinerant electrons in the NM. 
The strong coupling between magnons and electrons enhances the superconductive pairing. 
The dispersions of the conduction electrons and the magnons influence the pairing significantly. 
Choosing different combinations of materials and tuning the interface quality controls the superconductive gap. 

Extensive studies on the interplay between antiferromagnetic ordering and superconductivity have been conducted. 
Experiments have shown that the two phenomena can coexist in several different materials \cite{PhysRevLett.41.1133,PhysRevB.61.R14964} and even within the same electron bands \cite{PhysRevLett.60.615,PhysRevLett.75.1178,PhysRevB.56.11749}. 
Because many high-$T_C$ superconductors are created from antiferromagnetic insulators by doping \cite{RevModPhys.78.17}, their discovery led to a renewed interest in the relation between superconductivity and antiferromagnetism. 
Even more recently, superconductivity has been found to coexist with antiferromagnetism in iron pnictide superconductors \cite{PhysRevB.78.214515,PhysRevB.79.014506,NewJPhys2009Rotter,PhysRevLett.101.087001}. 

Theory predicts that magnons can mediate superconductivity in bulk antiferromagnets, with either p-wave or d-wave pairing symmetry \cite{PhysRevB.96.214409,JPSJ.63.1861}. 
There are also suggestions that magnons mediate superconductive pairing in iron pnictides \cite{KAR201818,WuJPC2011}. 

At topological insulator (TI)/FI interfaces, ferromagnetic magnons are predicted to mediate p-wave pairing of spin-momentum locked electrons, where the involved electrons can have equal momenta \cite{KagarianPRL2016}. 
For Bi/Ni bilayers, Ref.~\onlinecite{GongeSCIENCE2017} developed a similar model, but with a d-wave pairing, to explain their experimental findings of superconductivity. 
At TI/AFI interfaces, there are predictions that magnons mediate the pairing of spin-momentum locked electrons with either equal or antiparallel momenta \cite{hugdal2018magnon}. 

We consider pairing between spin-degenerate electrons in a metal. 
In Ref.~\onlinecite{PhysRevB.97.115401}, we showed that magnons in FIs can mediate the p-wave pairing of electrons with opposite momenta in FI/NM/FI tri-layers. 
In this paper, we replace the ferromagnetic insulators with antiferromagnetic insulators and consider AFI/NM/AFI tri-layers. 
Magnons in ferromagnets and antiferromagnets significantly differ, resulting in distinctive magnon-induced pairings.
For the AFI/NM/AFI system, we find d-wave pairing.

Our paper is organized as follows. In Sec.~\ref{sec:model}, we introduce the model describing the metallic layer, the antiferromagnetic layers, and the interaction between the layers. Sec.~\ref{sec:gap_equation} presents the resulting magnon-mediated electron-electron interaction, the gap equation, and its solution.
We conclude the paper in Sec.~\ref{sec:conclusions}.
Appendix \ref{app:mat} provides estimates for material parameters, and Appendix \ref{app:non-zero-momentum} considers an alternative superconducting pairing with a non-zero sum of the electron momenta and p-wave symmetry.
We will see that this pairing is suppressed compared to the d-wave pairing.

\section{Model}
\label{sec:model}

\begin{figure}[htb]
	\includegraphics[width=\columnwidth]{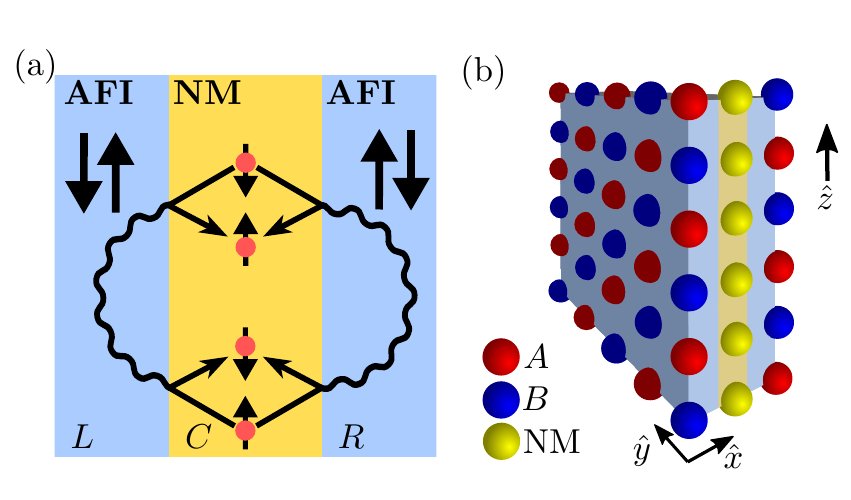}
	\caption{(Color online)
	Tri-layer system: normal metal sandwiched between two antiferromagnetic insulators. 
	(a) Electrons in the NM scatter at the interfaces, creating or annihilating a magnon. 
	This leads to an effective electron-electron interaction. 
	The spin of the electron is flipped in each scattering event. 
	(b) Three-monolayer lattice structure and coordinate axes $x$, $y$, and $z$. 
	}
	\label{fig:system}
\end{figure}

Our model consists of three monolayers: a NM sandwiched between two identical easy-axis AFIs, as shown in Fig.~\ref{fig:system}. 
We denote the left (right) AFI by $\Gamma = \L$ ($\R$) and the central NM by $\Gamma = \C$. 
We assume that all three layers have identical square lattices with lattice constant $d$, where node $i$ has the same in-plane position vector $\vec{r}_{i}$ in all layers $\R$, $\C$, and $\L$. 
We define the unit vectors $\hat{y}$ and $\hat{z}$ along the lattice vectors, and $\hat{x}$ is transverse to the monolayers. 
We characterize the spin directions with the coordinates $\chi$, $\upsilon$, and $\zeta$, where $\hat{\zeta}$ is parallel to the easy axis of the AFI. 
There are $N_y$ lattice nodes in the $y$ direction and $N_z$ lattice nodes in the $z$ direction. The total number of sites in the metal layer is $N = N_y N_z$. We use periodic boundary conditions along the $y$- and $z$-directions. 

We describe both AFIs using Heisenberg Hamiltonians with nearest-neighbor exchange interaction $J$ and easy-axis anisotropy $K_{\zeta}$, 
\begin{equation}
	H_{\AFI}^{\Gamma} = \frac{J}{\hbar^2} \sum_{ \nearnb{i}{j}}\vec{S}_{i}^{\Gamma} \cdot \vec{S}_{j}^{\Gamma} + \frac{K_{\zeta}}{\hbar^2} \sum_{i} \left( S_{i \zeta}^{\Gamma} \right)^2 \, .
\end{equation}
Here, $\hbar$ is the reduced Planck constant, $\vec{S}_{i}^{\L}$ ($\vec{S}_{i}^{\R}$) is the spin at node $i$ in the left (right) AFI, and $\nearnb{i}{j}$ is a pair of nearest-neighbor nodes. 
Each AFI is divided into two sublattices: $A$ and $B$. 
When the AFI is in its classical ground state, all the spins on sublattice $A$ ($B$) point along $\hat{\zeta}$ ($-\hat{\zeta}$). 
We assume that the matching nodes in the left and right AFIs are in opposite sublattices so that $\vec{S}_{i}^{\L} = - \vec{S}_{i}^{\R}$ in the classical ground state; see Fig.~\ref{fig:system} (b). 

\begin{figure}[ht]
	\includegraphics[width=\columnwidth]{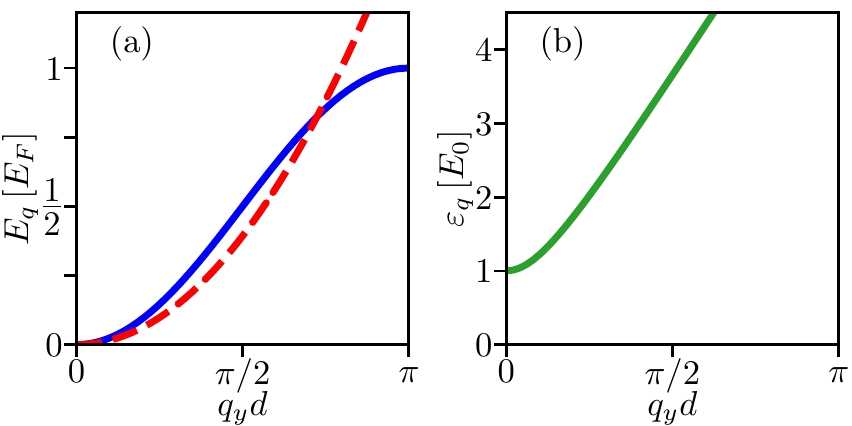}
	\caption{(Color online)
	Dispersion relations along $q_z = 0$ for (a) the conduction electrons and (b) the magnons, 
	assuming $J/K_{\zeta} = 10$ in the antiferromagnets. 
	The quadratic electron dispersion $E^{\Quad}_{\vec{q}}$ is the red dashed line, and
	the tight-binding dispersion $E^{\TB}_{\vec{q}}$ is the blue solid line. 
	}
	\label{fig:dispersion}
\end{figure}

For the electronic states, we consider two different models. 
The plane-wave states $c_{\vec{q},\sigma} = \sum_{j} \exp( i \vec{r}_{j} \cdot \vec{q} ) c_{j \sigma} /\sqrt{N} $ are eigenstates of both models, but the energy dispersions differ. 
In the first case, the energy dispersion follows from the tight-binding model ($\TB$). 
In the second case, we assume that the electron dispersion is quadratic ($\Quad$). 
The Hamiltonian of the tight-binding model is 
\begin{equation}
	H_{\TB} = - t \sum_{\sigma} \sum_{ \nearnb{i}{j} } c^{\dag}_{i \sigma} c_{j \sigma} \, ,
\end{equation}
where $c_{j \sigma}$ ($c^{\dag}_{j \sigma}$) annihilates (creates) a conduction electron with spin $\sigma$ along $\hat{\zeta}$ at node $j$. 
The plane-wave states are eigenstates of this Hamiltonian with the dispersion $E_{\vec{q}}^{\TB} = 2 t \left[ 2 - \cos(q_y d) - \cos(q_z d) \right]$. 
For the quadratic model ($\Quad$), we assume that the dispersion is $E^{\Quad}_{\vec{q}} = \hbar^2 \vec{q}^2 / ( 2m )$.
Here, $m$ is the effective electron mass. 
We assume half-filling in both models. 
The electron dispersion relations are illustrated in Fig.~\ref{fig:dispersion} (a). 

The spins in the AFIs couple to the conduction electrons via an interfacial exchange coupling $J_I$,
\begin{equation}
	H_{\Int} =  - \frac{J_I}{\hbar} \sum_{\sigma \sigma'} \sum_{j} \sum_{\Gamma=\L,\R} c^{\dag}_{j \sigma} \vecS{\sigma}_{\sigma \sigma'} c_{j \sigma'} \cdot \vec S_j^{\Gamma} \, . 
\end{equation}
Here, $\vecS{\sigma} = \hat{\chi}\sigma_x + \hat{\upsilon}\sigma_y + \hat{\zeta}\sigma_z$, and $\sigma_x$, $\sigma_y$, and $\sigma_z$ are the Pauli matrices. 

We perform a Holstein-Primakoff transformation, treating the sublattices $A$ and $B$ separately, and we define $S^{\Gamma}_{i \pm} = S^{\Gamma}_{i \chi} \pm i S^{\Gamma}_{i \upsilon}$. 
Assuming that the AFIs are close to their classical ground states, 
we find, for sublattice $A$, $S^{\Gamma}_{i +} = S^{\Gamma \dag}_{i -} = \hbar \sqrt{2s} a^{\Gamma}_{i} $ and $S^{\Gamma}_{i \zeta} = \hbar ( s - a^{\Gamma \dag}_{i} a^{\Gamma}_{i} )$, 
and, for $B$, $S^{\Gamma}_{i +} = S^{\Gamma \dag}_{i -} = \hbar \sqrt{2s} b^{\Gamma \dag}_{i}$ and $S^{\Gamma}_{i \zeta} = \hbar ( b^{\Gamma \dag}_{i} b^{\Gamma}_{i} - s )$. 
Using Fourier- and Bogoliubov transformations, we obtain the magnon eigenstates
\begin{equation}
		a^{\Gamma}_{\vec{k}} = \sqrt{\frac{2}{N}} \left( \sum_{i \in A} u_{\vec{k}} e^{-i \vec{k} \vec{r}_{i}} a_{i}^{\Gamma} - \sum_{i \in B} v_{\vec{k}} e^{i  \vec{k} \vec{r}_{i }} b^{\Gamma \dag}_{i} \right) \, .
\end{equation}
The expression for $b^{\Gamma}_{\vec{k}}$ is found by exchanging $a$ and $b$. 
The Bogoliubov constants $u_{\vec{k}}$ and $v_{\vec{k}}$ satisfy $u_{\vec{k}}^2 - v_{\vec{k}}^2 = 1$.

We assume the anisotropy $K_{\zeta}$ is substantially smaller than the exchange $J$ so that \cite{PhysRevB.95.144408} $u_{\vec{k}} \approx - v_{\vec{k}} \approx \sqrt{ \varepsilon_J / \varepsilon_{\vec{k}} }/ \sqrt[4]{2} \gg 1$. 
Because the dominant contribution to the superconducting gap is expected to come from the long-wavelength magnons \cite{PhysRevB.97.115401}, we will use this so-called exchange approximation throughout. In the long-wavelength limit, the magnon dispersion is $\varepsilon_{\vec{k}} = 2 s \sqrt{ 2 J \left( 2 K_{\zeta} + J \vec{k}^2 d^2 \right) }$. 
In terms of the magnon gap $\varepsilon_0 = 4 s \sqrt{ J K_{\zeta} }$ and the exchange energy scale $\varepsilon_J = 2 \sqrt{2} J s$, the dispersion is $\varepsilon_{\vec{k}} = \sqrt{\varepsilon_0^2 + \varepsilon_J^2 \vec{k}^2 d^2}$.

\begin{figure}[htb]
	\includegraphics[width=\columnwidth]{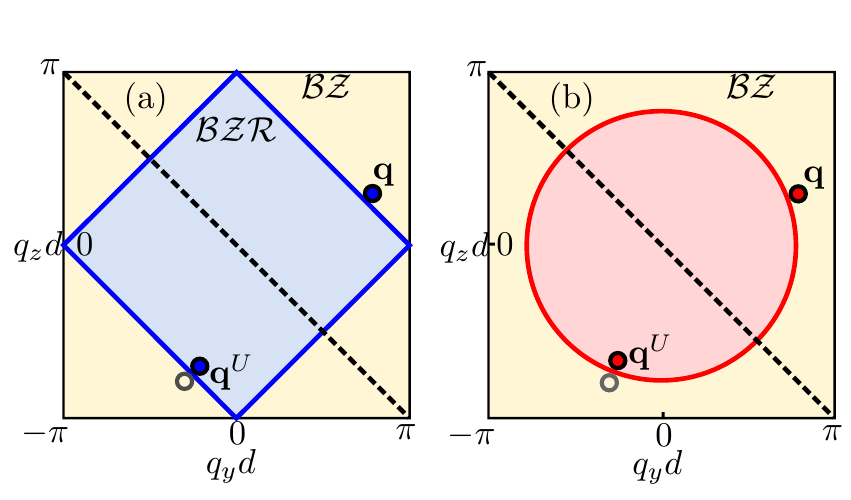}
	\caption{(Color online)
	Fermi surfaces for (a) the tight-binding model (blue) and (b) the quadratic model (red). 
	The Brillouin zone of the conduction electrons ($\BZ$) is shown in yellow. 
	The reduced (magnon) Brillouin zone ($\BZR$) corresponds to the interior of the Fermi surface of the tight-binding model (light blue). 
	The Umklapp momentum $\vec{q}^{\U}$ is related to $\vec{q}$ by a reflection across the diagonal of the $\BZ$ (dashed line)
	and a subsequent reflection in the Fermi surface. 
	}
	\label{fig:Brillouin}
\end{figure}

The momenta ($\vec{q}$) of the conduction electrons reside in the Brillouin zone, $\BZ$, of the lattice of the NM. 
By contrast, the magnon momenta ($\vec{k}$) are defined in the reduced Brillouin zone of the sublattices, $\BZR$; see Fig.~\ref{fig:Brillouin} (a). At half-filling, the $\BZR$ matches the interior of the Fermi surface of the tight-binding model. 

We disregard terms of second order in the magnon operators from $H_{\Int}$. 
Then, the total Hamiltonian $H = H_{\AFI}^{\L} + H_{\AFI}^{\R} + H_{\NM} + H_{\Int}$ is given by \cite{PhysRevB.95.144408} 
\begin{align}
	H &= \sum_{\Gamma}\sum_{\vec{k}} \varepsilon_{\vec{k}} \left( a_{\vec{k}}^{\Gamma \dag} a_{\vec{k}}^{\Gamma} + b_{\vec{k}}^{\Gamma \dag} b_{\vec{k}}^{\Gamma} \right) + \sum_{\vec{q} \sigma}E_{\vec{q}}c_{\vec{q},\sigma}^\dag c_{\vec{q},\sigma} \\
    &+ \sum_{\Gamma} \sum_{\vec{k} \vec{q}} \br{V}_{\vec{k}} \left( a_{\vec{k}}^{\Gamma} c^{\dag}_{\vec{q}^{\U},\downarrow} c_{\vec{q}-\vec{k}, \uparrow} + b_{\vec{k}}^{\Gamma}c^{\dag}_{\vec{q}^{\U},\uparrow}c_{\vec{q}-\vec{k}, \downarrow} + \text{h.c.} \right) \notag ,
\end{align}
where
\begin{equation}
\label{eq:Umklapp}
 \vec q^\U = \vec q + \vec q_{\AF}\text{ with }\vec q_{\AF} = \left(\hat{y} + \hat{z} \right) \pi/d
\end{equation}
is the Umklapp momentum of $\vec q$ and $\br{V}_{\vec{k}} = - \sqrt{s/2N} J_I u_{\vec{k}}$. Importantly, the interfacial coupling $\br{V}_{\vec{k}}$ is enhanced by the Bogoliubov constants relative to the magnon-electron coupling in ferromagnets \cite{PhysRevB.97.115401}. 
To leading order in the exchange approximation, the conduction electrons only interact with magnons through Umklapp scattering. 
In contrast to NM-AFI bilayers, the contribution from the normal channel is negligible because the static spin-dependent potentials from the two AFIs compensate each other almost completely. 

In the electronic tight-binding model at half-filling, the Umklapp process $\vec{q} \to \vec{q}^{\U}$ can be split into two steps; see Fig.~\ref{fig:Brillouin} (a). 
First, there is a reflection across one of the diagonals of the full Brillouin zone ($\BZ$). 
Second, there is a reflection across the Fermi surface. 
The second reflection occurs at the surface parallel to the diagonal of the first reflection. For initial states on the Fermi surface, an Umklapp process takes a state $\vec k$ to another state $\vec k^\U$ that is also on the Fermi surface. 

Next, we consider the approximate model with quadratic electron dispersion $E^{\Quad}_{\vec{q}}$. To retain the main physics of the tight-binding model, we introduce a modified Umklapp momentum $\vec{q}^{\MU}$ that contains two analogous consecutive reflections. The first reflection is across one of the diagonals of the $\BZ$. The second reflection is across the circular Fermi surface corresponding to the quadratic electron dispersion; see Fig.~\ref{fig:Brillouin} (b). 
The definition of $\vec{q}^{\MU}$ depends on the choice of the diagonal where the first reflection occurs.
We remove this ambiguity by requesting that $\sign(q_y^{\MU})\sign(q_z^{\MU}) = \sign(q_y)\sign(q_z)$.
However, for the symmetries of the superconducting gap that we consider in the following section, all choices for the first reflection lead to the same results.

In Sec.~\ref{sec:quadratic}, we will see that the simplifications associated with the rotational symmetry of the quadratic dispersion together with the modified Umklapp process allow for exploration of a large range of parameters as the angular dependence of the gap can be treated analytically. 

\section{Gap Equation}
\label{sec:gap_equation}

Integrating over all the magnons, we find the magnon-mediated electron-electron interaction
\begin{equation}
  H_{\inte} = \sum_{\vec{q}\vec{p}\vec k} \tilde{V}_{\vec{k}\vec{q}\vec{p}} c^\dag_{ \vec{p}^\U ,\downarrow} c^\dag_{\vec{q}-\vec{k} \uparrow} c_{\vec{p}-\vec{k} \uparrow} c_{ \vec{q}^{\U}, \downarrow} \, .
  \label{eq:Hint}
\end{equation}
The interaction of Eq.~(\ref{eq:Hint}) influences all the electrons. 
We focus on the possible formation of Cooper pairs. 
We consider the scenario, whereby the essential terms in Eq.\ (\ref{eq:Hint}) satisfy $\vec{p} = -\vec{q}^\U + \vec{k}$. 
Then, the two electrons forming a pair have opposite momenta as in the BCS theory. 
Another possibility will be discussed in Appendix \ref{app:non-zero-momentum}.

The effective interaction simplifies to
\begin{equation}
	H = \sum_{\vec{q} \vec{p}} V_{\tld{\vec{q}}, \vec{p}} c_{\tld{\vec{q}} \downarrow}^{\dag} c_{-\vec{q} \uparrow}^{\dag} c_{-\vec{p}\uparrow} c_{\tld{\vec{p}} \downarrow} \, ,
\end{equation}
where the effective coupling is 
\begin{equation}
	V_{\vec{q}, \vec{p}} = \frac{4 J_I^2 J s^2}{N_y N_z} \frac{ \theta_{ \vec{q}^{\U} + \vec{p} } }{\varepsilon_{\vec{q}^{\U}{+}\vec{p}}^2 - (E_{\vec{q}} - E_{\vec{p}})^2} \, .
	\label{eq:gap.effectiveinteraction}
\end{equation}
Here, we have used the step function $\theta$, where $\theta_{ \vec{q} } = 1$ when $\vec{q}$ is inside the $\BZR$ and $\theta_{ \vec{q} } = 0 $ otherwise.

We define a spin-singlet gap function 
\begin{equation}
	\Delta_{\vec{q} } = \sum_{ \vec{p} } V_{ \vec{q} \vec{p} } \langle c_{-\vec{p}\uparrow} c_{ \vec{p} \downarrow} - c_{-\vec{p}\downarrow} c_{\vec{p} \uparrow} \rangle \, .
	\label{eq:gapopp}
\end{equation}
The corresponding gap equations is
\begin{equation}
 \Delta_{\vec{q} } = - \sum_{\vec{q}'} V_{\tld{\vec{q}}, \vec{q}'} \frac{\Delta_{\vec{q}'}}{2\tilde{E}_{\vec{q}'}} \tanh\left(\frac{\tilde{E}_{\vec{q}'}}{2k_B T}\right) \, ,
 \label{eq:gap}
\end{equation}
where $\tilde{E}_{\vec{q}} = \sqrt{(E_{\vec{q}}-E_F)^2+\abs{\Delta_{\vec{q}}}^2}$, $k_B$ is the Boltzmann constant, and $T$ is the conduction-electron temperature. 

In order to determine the symmetry of the gap function, we consider the case where the dominant part of $V_{\tld{\vec{q}}, \vec{q}'}$ in Eq.~\eqref{eq:gap} comes from the long-wavelength magnons $\vec{q}^{\U}{+}\vec{q}' \approx \vec{0}$, as in Ref.~\onlinecite{PhysRevB.97.115401}. 
Then, we expect that $\Delta_{ -\vec{q}^{\U} } \approx - \Delta_{ \vec{q} }$, where the minus follows from comparing the sign in Eq.~\eqref{eq:gap} with the BCS theory or Ref.~\onlinecite{PhysRevB.97.115401}. 
In the tight-binding model, these relations are satisfied if the gap function $\Delta$ is of d-wave symmetry, i.e., it satisfies
\begin{equation}
    \Delta_{ (q_y,q_z) } = -\Delta_{ (-q_z,q_y) } = \Delta_{ (-q_y,q_z) } = \Delta_{ (q_y,-q_z) }. \label{eq:gapsymmopp} 
\end{equation}
We assume that the superconducting gap has the same symmetry in the quadratic model. 

To solve the gap equation, we replace the sum over momenta with integrals over the energy $E = E_{\vec{q}}$ and the angle $\varphi$, where $\vec{q} = {q} \left[\sin(\varphi),\cos(\varphi) \right]$. 
We assume that the dominant contribution to the effective coupling $V_{\tld{\vec{q}}, \vec{q}'}$ in Eq.~\eqref{eq:gap} stems from the regions where $\vec{q}^{\U}{+}\vec{q}'$ lies within the reduced Brillouin zone $\BZR$. 
We therefore set $\theta_{\vec{q}^{\U} + \vec{p}} = 1$ for all $\vec{q}$ and $\vec{p}$ in Eq.~\eqref{eq:gap.effectiveinteraction}. 
We then introduce dimensionless variables in terms of the magnon gap, $\varepsilon_0$, such that $\delta = \Delta/\varepsilon_0$, $\tau = k_B T/\varepsilon_0$, $x = (E-E_F)/\varepsilon_0$, $\tilde{x} = \tilde{E}/\varepsilon_0$, and $\epsilon = \varepsilon/\varepsilon_0$.
The gap $\delta=(x,\varphi)$ has to satisfy the self-consistent equation
\begin{equation}
    \delta(x,\varphi) = -\tilde\alpha\!\! \int\limits_{-x_B}^{x_B}\!\!\!\!dx' \! \! \int\limits_0^{2\pi}\!\!d\varphi'\,
                        \frac{\delta(x',\varphi')v(x,x',\varphi,\varphi')}{\tilde x'} \tanh\left[\frac{\tilde x'}{2\tau}\right]
\label{eq:gap_eeq2D}
\end{equation}
with the dimensionless coupling strength $\tilde\alpha=J_I^2 s \varepsilon_J/ ( 2 \pi \sqrt{2} E_F \varepsilon_0^2 )$, $\tilde x' = \sqrt{(x')^2+|\delta(x',\varphi')|^2}$, and
\begin{equation}
 v(x,x',\varphi,\varphi') = \frac{1}{1+\varepsilon_J^2 |\vec k^U\!\!+\vec k'|^2d^2/ \varepsilon_0^2  - (x{-}x')^2} \propto V_{\vec k,\vec k'}
 \label{eq:pot_2D}
\end{equation}
where we approximate $\vec k$ by $\vec k = k_F(\hat{y} \sin\varphi + \hat{z} \cos\varphi)$ and $k_F = \sqrt{2\pi}/d$.
The dependence of $\vec k$ on $x$ is disregarded since $x\ll E_F/\varepsilon_0$.
This means that the magnon energy depends solely on the angles: $\varepsilon=\varepsilon(\varphi,\varphi')$.
We restrict the energy $x'$ to an interval $[-x_{\B},x_{\B}]$, where $x_{\B} > 1$ is chosen such that $\abs{\delta(x',\varphi')} \ll \max_\varphi\abs{\delta(0,\varphi)}$ for all $x'$ outside the interval. 

The remainder of this section is organized as follows. In Sec.~\ref{sec:quadratic}, we solve the gap equation for a simplified model.
In this model, we assume a quadratic electron dispersion together with the modified Umklapp momentum, $\vec q^{\MU}$, introduced at the end of Sec.~\ref{sec:model}.
We explore the dependence of the superconducting gap on the coupling strength and on the temperature in the exchange limit where the magnon gap is smaller than the exchange energy, $\varepsilon_0/\varepsilon_J\ll1$.
The purpose of obtaining these results is to give a basic understanding of the physics.

In Sec.~\ref{sec:2D}, we and solve the gap equation numerically for the actual Umklapp relation from Eq.~(\ref{eq:Umklapp}) and the quadratic dispersion. 
Sec.~\ref{sec:tb} discusses differences in the tight-binding model compared to the calculations with the quadratic dispersion.

Finally, in Sec.~\ref{sec:analysis}, we will analyze the differences between the simplified model (Sec.~\ref{sec:quadratic}), the quadratic dispersion model (Sec.~\ref{sec:2D}), and the tight-binding model (Sec.~\ref{sec:tb}).

\subsection{Simplified model: quadratic electron dispersion with modified Umklapp relation}
\label{sec:quadratic}
Using the modified Umklapp relation is a great simplification because we can use the rotational symmetry.
The gap equation has a d-wave solution $\delta(x,\varphi)$ satisfying Eq.~\eqref{eq:gapsymmopp}. 
At the critical temperature, $\tau=\tau_c$, where the gap approaches zero, this state takes the form $\delta(x,\varphi) = f(x) \cos(2 \varphi) $, where $f$ satisfies 
\begin{equation}
 f(x) = \alpha \! \! \! \int\limits_{-x_{\B}}^{x_{\B}} \! \! dx' \frac{ V(x{-}x') f(x')}{ \sqrt{x^{\prime 2}{+}f(x')^2} } \tanh \! \! \left[\frac{ \sqrt{x^{\prime 2}{+}f(x')^2} }{2\tau}\right] \! .
	\label{eq:gap.quadratic}
\end{equation}
Here, $\alpha =  \tilde\alpha \sqrt{\pi/2} \cdot ( \varepsilon_0 / \varepsilon_J ) $ is the coupling constant, the effective potential is $V(y) \approx -C_{V} + 1/\sqrt{1 - y^2 }$, and the constant $C_{V}=\sqrt{2}\varepsilon_0/(\sqrt{\pi}\varepsilon_J)$, which we will set to zero in the numerical calculations.

Note that for $\tau<\tau_c$, $\tilde x'=\sqrt{(x')^2+|\delta(x',\varphi')|^2}$ depends in general on $\varphi'$.
Consequently, the integration over the angle $\varphi'$ cannot be separated from the integration over $x'$ as was done for the derivation of Eq.~(\ref{eq:gap.quadratic}). 
Thus, for temperatures below the critical temperature, using Eq.~(\ref{eq:gap.quadratic}) represents a simplifying assumption compared to solving Eq.~(\ref{eq:gap_eeq2D}) for $\delta(x',\varphi')$. 
However, the solution $f(x')$ to Eq.~(\ref{eq:gap.quadratic}) is approximately equal to the maximum amplitude of the d-wave gap for a given energy, $\max_{\phi'} \{ \delta(x',\varphi') \}$. 
Also, since Eq.~(\ref{eq:gap.quadratic}) is valid near the critical temperature, we can use it to calculate the critical temperature itself.
The p-wave gap function of Appendix \ref{app:non-zero-momentum} satisfies Eq.~(\ref{eq:gap.quadratic}) at all temperatures, so the results are also valid for this pairing. 

\begin{figure}[htp]
	\includegraphics[width=\columnwidth]{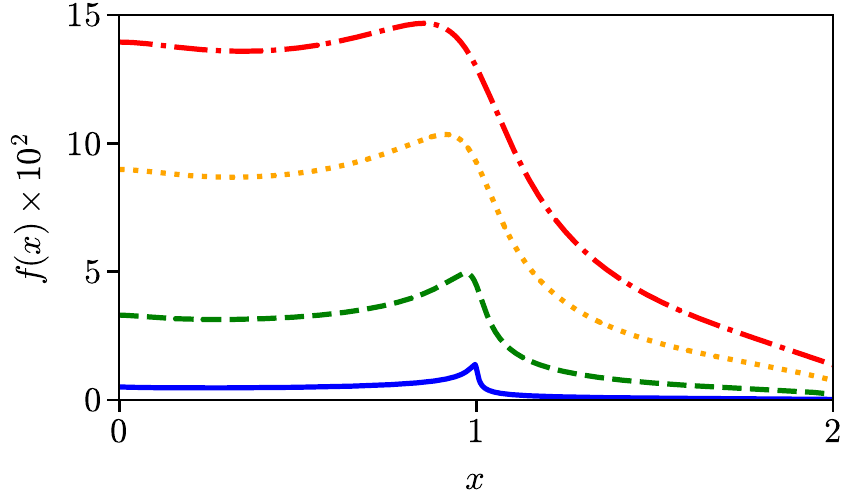}
	\caption{ (Color online)
	Numerical results for the energy dependence of the gap function $f(x)$ according to Eq.~\eqref{eq:gap.quadratic}
	at zero temperature ($\tau=0$),  found by iterations starting with a Gaussian. 
	The small constant $C_{V}$ was approximated as vanishing, $C_{V} = 0$. 
	We consider four different values of the dimensionless coupling constant $\alpha=0.07$ (blue solid line),
	                                                                         $\alpha=0.1$ (green dashed line),
	                                                                         $\alpha=0.13$ (orange dotted line), and
	                                                                         $\alpha=0.17$ (red dash-dot line).
	}
	\label{fig:quadratic_gap}
\end{figure}
We solve the 1D gap equation \eqref{eq:gap.quadratic} numerically by iteration. 
$f$ is symmetric about the Fermi surface: $f(-x)=f(x)$. 
Fig. \ref{fig:quadratic_gap} shows the solutions of Eq.~\eqref{eq:gap.quadratic} for different coupling constants $\alpha$ at zero temperature. 
We find a relatively constant behavior around $x=0$ and, for small $\alpha$, a pronounced peak at $\abs{x}\approx1$.

\begin{figure}[htb]
	\includegraphics[width=\columnwidth]{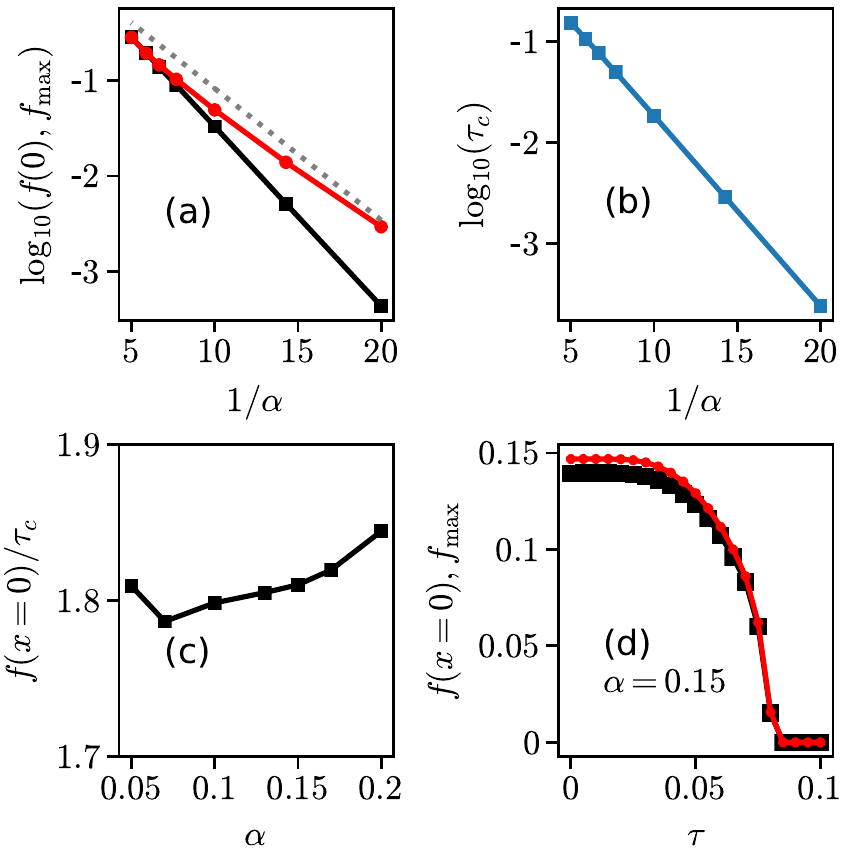}
	\caption{(Color online)
	Numerical results for $\alpha$ and for the temperature dependence of the gap function defined by Eq.~\eqref{eq:gap.quadratic}
	for the potential with a constant $C_{V}$ again set to be zero. 
	(a) Semi-logarithmic plot of $f(x{=}0)$ (black squares) and $f_{\rm max}=\max_x f(x)$ (red circles) for $\tau=0$ in the dependence of $1/\alpha$.
	    The gray dotted line refers to the BCS-like consideration of a constant potential $V(x,x')=V_c$ within $|x|,|x'|<1$,
	    which results in $f(x)=2\exp(-1/(\alpha\pi))$ for $|x|<1$.
	(b) Semi-logarithmic plot of the dimensionless critical temperature $\tau_c$ as a function of the coupling $1/\alpha$. 
	(c) Ratio of $f(x{=}0)$ at $\tau=0$ to the critical temperature $\tau_c$ as a function of $\alpha$. 
	(d) Temperature dependence of $f(x{=}0)$ (black squares) and $f_{\rm max}$ (red circles) for $\alpha = 0.15$. 
	}
	\label{fig:quadratic}
\end{figure}

We compare the $\alpha$ dependence of $f_{\rm max} = \max_{x}f(x)$ and $f(0)$ to the standard BCS result $f\sim\exp(-1/\alpha)$; see Fig.~\ref{fig:quadratic}~(a). 
The BCS result was derived for a potential $V(x,x')$ which is constant $V(x,x')= V_c$ if $|x|,|x'|<1$ and $0$ otherwise, for $V_c  = \pi/2=\int_{-1}^1 dy V(y)/2$. 
The $\alpha$ dependence of the critical temperature $\tau_c$ is comparable to the one of $f(x{=}0)$; see Fig.~\ref{fig:quadratic}~(b).
The ratio $f(x{=}0)/\tau_c$ is slightly higher in our model than in standard BCS theory, where the ratio is approximately $1.76$; see Fig.~\ref{fig:quadratic}~(c).
Note that the angle dependence is already integrated out in Eq.~(\ref{eq:gap.quadratic}), and a constant potential would result in the $1.76$ ratio.
When we vary the temperature $\tau$, $f_{\rm max}$ and $f(x{=}0)$ both vanish at $\tau_c$, as expected. 
As we see from Fig.~\ref{fig:quadratic}~(d), $f_{\rm max}$ and $f(x{=}0)$ show similar $\tau$ dependencies.

In making the model dimensionless, the magnon gap $\varepsilon_0$ is a natural choice of energy scale. 
In the resulting gap equation, the coupling $\alpha$ is inversely proportional to $\varepsilon_0$. 
As we observed in Fig.~\ref{fig:quadratic}, $\tau_c$ scales similarly to $\exp(-1/\alpha)$. 
Therefore, $T_c$ might increase by reducing the magnon gap $\varepsilon_0$. 
However, if we increase $\alpha$, the system eventually enters a regime where higher order effects will have to be considered. 
For FI/NM/FI tri-layers, the exchange energy scale $\varepsilon_J$ plays the same role as $\varepsilon_0$ for AFI/NM/AFI tri-layers \cite{PhysRevB.97.115401}. 
Because $\varepsilon_J$ is typically larger than $\varepsilon_0$, $T_c$ should in many cases be higher for AFI/NM/AFI tri-layers than for FI/NM/FI tri-layers, assuming that the coupling $J_I$ is the same. 
However, the strong-coupling regime may set in at lower values of $J_I$ for AFIs than FIs since the coupling constant $\alpha$ is typically larger for AFIs. 

We estimate $\varepsilon_0$ and $\alpha$ for a Mn$\text{F}_2$-Au-Mn$\text{F}_2$ tri-layer in Appendix \ref{app:mat}. 
We find $\varepsilon_0/k_B = 13$~K and the range of values $[0.02\text{--}0.18]$ for $\alpha$. 
For the simplified model, the corresponding critical temperatures are up to the order of one Kelvin. 
We assume that $J_I$ is similar in magnitude for AFI/NM interfaces as for FI/NM interfaces. 
Similar assumptions have been made in earlier work \cite{PhysRevLett.113.057601,PhysRevB.90.094408}. 
In our model, $J_I$ represents the strength of the interfacial electron-magnon coupling. 
Spin transport across AFI/NM interfaces has been measured in several experiments \cite{PhysRevLett.115.266601,PhysRevLett.116.097204,JAppPhys.118.233907}. 
The spin transport between an FI and a NM can be enhanced by inserting an AFI in between, indicating that the coupling at AFI/NM interfaces is as strong as compared to FI/NM interfaces \cite{PhysRevLett.113.097202}. 

\subsection{Quadratic electron dispersion with the actual Umklapp relation}
\label{sec:2D}
Now we consider the quadratic dispersion relation together with the actual Umklapp relation and solve Eq.~(\ref{eq:gap_eeq2D}) numerically.
\begin{figure}[htp]
    \centering
    \includegraphics{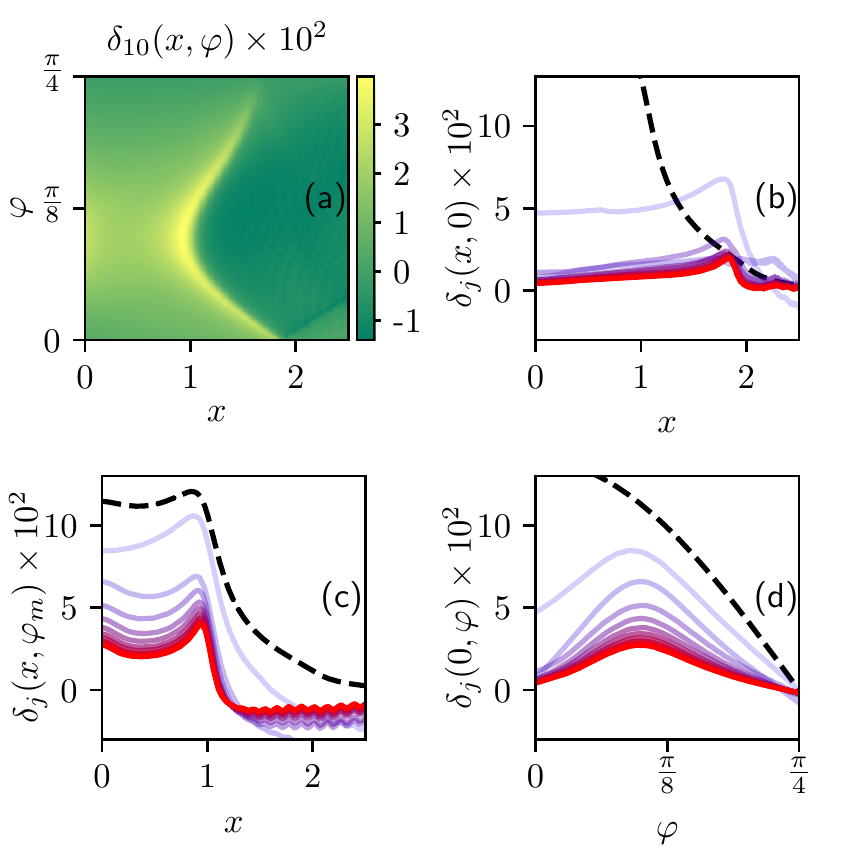}
    \caption{(Color online)
    Iterative solution of the gap equation Eq.~(\ref{eq:gap_eeq2D}) for the gap $\delta$ as a function of dimensionless energy $x$ and angle $\varphi$ at zero temperature and $\alpha=0.15$. The initial guess is $\delta_0(x,\varphi)=f(x) \cos(2\varphi)$, where $f(x)$ is the solution of Eq.~(\ref{eq:gap.quadratic}) obtained previously.
    (a) Gap after ten iterations, $\delta_{10}(x,\varphi)$. If the gap $\delta(x,\varphi)$ is known for $\varphi\in[0,\pi/4)$ and $x>0$, its values at all other points in k space follow from symmetry.
    (b) Gap as a function of $x$ for $\varphi=0$.
    (c) Gap as a function of $x$ at the angle $\varphi_m=\arcsin(\pi/2{-}1)/2\approx0.3$, where the Fermi surface and the boundary of the $\BZR$ intersect.
    (d) Gap as a function of $\varphi$ for $x=0$.
    In (b-d), the iterations are $j=0$ (black dashed line), and then, $j=1,\ldots,10$, shown in light blue (light gray) to red (darker gray).
    }
    \label{fig:2D}
\end{figure}

In Fig.~\ref{fig:2D}, we present iterative results for Eq.~(\ref{eq:gap_eeq2D}) for $\alpha=0.15$ at zero temperature.
The initial guess for the iterations is the d-wave gap function $\delta_0(x,\varphi) = f(x)\cos(2\varphi)$, where $f(x)$ is the solution of Eq.~(\ref{eq:gap.quadratic}) with $C_V = 0$.

As we see in Fig.~\ref{fig:2D}(b-d), the gap function converges after a few iterations.
The resulting function is smaller compared to the initial guess.
We find the highest values at the angle $\varphi_m = \arcsin(\pi/2{-}1)/2$, where the Fermi surface intersects with the boundary of the $\BZR$.
At this point in $k$ space, the modified Umklapp relation $Q^{\MU}$ used previously is equal to the actual Umklapp relation $Q^{\U}$.

The gap function $\delta(x,\varphi)$ does not have the $\cos(2\varphi)$ dependence on $\varphi$; see Fig.~\ref{fig:2D}(d).
The reason is the difference between the actual Umklapp relation and the simplified one used previously.
We anticipate a similar behavior at finite temperature, as the simplified Umklapp relation remains only accurate at $\varphi=\varphi_m$.

For the critical temperature, we find numerically $\tau_c=0.012$ and a ratio $\delta(x{=}0,\varphi{=}\varphi_m)/\tau_c=2.1$. 
The ratio is slightly larger than the results in Fig.~\ref{fig:quadratic_gap} (c). 

To summarize the numerical results for the non-simplified model with quadratic dispersion relation: a solution of the gap equation with opposite-momentum pairing of d-wave type exists.

\subsection{Specifications of the gap equations in the tight-binding model}
\label{sec:tb}
We noticed that at half filling, the $\BZR$ is identical to the Fermi surface of the tight-binding model for the electrons.
This means that for the tight-binding model, the Um\-klapp process relates one point at the Fermi surface to another one at the Fermi surface. 
This indicates that the pairing mechanism is efficient at the Fermi energy similar to the simplified model considered in Sec.~\ref{sec:quadratic}.
However, there are differences in the tight-binding model compared to the simplified model that can have a significant impact on the superconductivity.
In contrast to the circular Fermi surface of the quadratic dispersion, the tight-binding half-filling Fermi surface touches the boundary of the $\BZ$, implying additional boundary conditions.
A d-wave gap symmetric gap function satisfies the additional boundary conditions in the sense that it is continuous at the edges of the $\BZ$.
Thus, we conclude that the d-wave gap can be the dominant contribution to superconductivity; compare with Appendix B.

We have observed that the d-wave gap is robust in the two models considered. 
We believe that it will remain robust even when including the full electronic tight-binding dispersion. 
The increased density of states near the corners of the Fermi surface may enhance the amplitude of the superconducting gap and the critical temperature. 
However, $\varepsilon_0$ remains the natural choice of energy scale. 
As we saw in section \ref{sec:2D}, the scale of the superconducting gap can be up to the order $\varepsilon_0/10$ for the quadratic dispersion. 
The gap in the tight-binding model is expected to be of the same order or higher. 
If the gap in the tight-binding model is of the order of $\varepsilon_0$ or larger, higher-order effects may have to be included. 
These effects, together with the tight-binding dispersion, add considerable complexity to the problem, which is beyond the scope of this initial work.

\subsection{Analysis of the solution of the gap equation}
\label{sec:analysis}

As we see from comparing Secs.~\ref{sec:quadratic} and \ref{sec:2D}, the pairing symmetry depends on the details of the electron dispersion and its interplay with the Um\-klapp process, which we will analyze in the following.

Um\-klapp scattering dominates the electron-magnon scattering in the scenario that we consider here. This situation differs when the antiferromagnetic sublattices couple unequally to the metal layer; see Ref.~\onlinecite{Erlandsen2019}.

From Sec.~\ref{sec:2D}, we see that the opposite-momenta d-wave gap has the highest amplitude where the Fermi surface intersects with the $\BZR$.
We assume that the same is the case for all electron dispersion relations.

The energy scale of the superconducting pairing is given by the magnon gap $\varepsilon_0$.
This differs from the results obtained for FI/NM/FI systems, where the relevant energy scale is the exchange energy between the spins in the FI layers \cite{PhysRevB.97.115401}.
In the AFI/NM/AFI system, the exchange energy $\varepsilon_J$ drops out of the gap equation completely for the simplest case, as $\alpha$ in Eq.~(\ref{eq:gap.quadratic}) does not depend on it at all.
The reason for this is an interplay of the Bogoliubov coefficients and the angular dependence of the gap equation together with the fact that $\varepsilon_J/\varepsilon_0\ll1$.

A further difference between the AFI/NM/AFI system with respect to the FI/NM/FI tri-layer is in the dependence of the size of the superconducting gap and the critical temperature on the dimensionless coupling constant $\alpha$.
Note that $\alpha$ is quadratic in the interfacial coupling $J_I$ ($\alpha\sim J_I^2$).
For the FI/NM/FI system, we found a dependence close to $f(x)\sim \alpha^2$ \cite{PhysRevB.97.115401}; we find here a behavior similar to the constant-potential result $f(x)\sim\exp(-1/\alpha)$.
The origin of this difference lies in the fact that here the width of the gap $f(x)$ is given approximately by $2\varepsilon_0$, whereas for FI/NM/FI, it was dependent on $\alpha$.

\section{Conclusions}
\label{sec:conclusions}

In conclusion, we predict that magnons mediate superconductivity in anti\-ferro\-mag\-netic insulator-metal-anti\-ferro\-mag\-netic insulator tri-layers.  
The exchange interaction at the antiferromagnet insulator-normal metal interfaces couples the electrons to the magnons. 
The influence of the interaction is, therefore, most potent when the metal is thin. 
We find superconducting d-wave pairing of electrons with opposite momenta.
The d-wave pairing dominates over p-wave finite momentum pairing, considered in Appendix \ref{app:non-zero-momentum}.
We find that the critical temperature is closely related to the magnon gap in the antiferromagnets. We estimate the critical temperature for a combination of Mn$\text{F}_2$ and Au to be on the order of Kelvin. 

\begin{acknowledgments}
We thank Asle Sudb\o{}, Eirik Erlandsen, and Akashdeep Kamra for useful discussions.
This work was partially supported by the European Research Council via Advanced Grant No. 669442 ``Insulatronics'', the Research Council of Norway through its Centers of Excellence funding
scheme, project number 262633, ”QuSpin”,
and by the Deutsche Forschungsgemeinschaft (DFG) under project number 417034116.
\end{acknowledgments}

\appendix

\section{Material Parameters}
\label{app:mat}

As a candidate AFI, we consider a (111)-layer of Mn$\text{F}_2$. 
Mn$\text{F}_2$ is an AFI with a large uniaxial anisotropy. 
The $s = 5/2$ Mn-ions in the (111)-layer form a square lattice with a lattice constant of $3.82$~\r{A} \cite{JAUCH1983907}. 
Based on measurements of the spin-wave dispersion of Mn$\text{F}_2$, we find $J/k_B = 4.1$~K and $K_{\zeta}/k_B = 0.39$~K \cite{JAppPhys.35.998}. 

For the normal metal, we consider a monolayer of gold with the same lattice structure as Mn$\text{F}_2$. 
We estimate the effective mass using $m=2\pi g_{\Sh}\hbar^2/E^{\B}_F$, where $g_{\Sh} = 12$~$\text{nm}^{-2}$ \cite{TserkovnyakRMP2005} is the Sharvin conductance and $E^{\B}_{F} = 5.5$~eV \cite{ashcroft1976} is the bulk Fermi energy.
We use the quadratic model and the assumption of half-filling to estimate the Fermi energy of the monolayer: $E_{F} = 1.6$~eV. 

As explained in Sec.~\ref{sec:quadratic}, we assume that the interfacial exchange coupling $J_I$ is similar in magnitude at AFI/NM interfaces compared to FI/NM interfaces. 
We therefore estimate $J_I$ using experimental values for the FI/SC interfaces, where the superconductor (SC) is either aluminum or vanadium. 
Estimates for the exchange coupling
\footnote{The exchange coupling as given in Ref.~\cite{Miao_et_al_NatComm2013} corresponds to $J_I/2$. }
within the range $[10\text{--}30]$~meV have been given for several such interfaces \cite{Tkaczyk_thesis,Roesler_et_al_1994,Miao_et_al_NatComm2013}.
Using $\alpha = J_I^2 / ( 16 E_F \sqrt{\pi JK} )$, we find a range of values $[0.02\text{--}0.18]$ for $\alpha$.

\section{Non-zero-momentum pairing}
\label{app:non-zero-momentum}
In this appendix, we consider an alternative type of superconducting pairing, electron pairs with nonzero total momentum, where the important terms in Eq.\ (\ref{eq:Hint}) are those with $\vec{p} = - \vec{q} + \vec{k}$; then, the Hamiltonian reduces to
\begin{equation}
	H = \sum_{\vec{q} \vec{p}} V_{\tld{\vec{q}}^{\U}, \vec{p}} c_{\tld{\vec{q}}^{\U} \downarrow}^{\dag} c_{-\vec{q} \uparrow}^{\dag} c_{-\vec{p}\uparrow} c_{\tld{\vec{p}}^{\U} \downarrow} \, ,
\end{equation}
which means that the sum of the momenta of the paired electrons is $\vec{q}_{\AF}$.
Here, the gap has a p-wave symmetry. 
Electron pairs with a total momentum of $\vec{q}_{\AF}$ were proposed in Ref.~\onlinecite{PhysRevLett.46.614} for bulk s-wave superconductors with antiferromagnetic order.

For the non-zero momentum pairing, the gap can be defined in two possible ways:
\begin{align}
	\Delta_{\vec{q} } &= \label{eq:gapnz} \\
	&\sum_{ \vec{p} } \! V_{ \vec{q}^{\U} \vec{p} } \! \langle c_{{-}\vec{p}\uparrow} c_{\vec{p}^{\U} \downarrow} {\pm} c_{{-}\vec{p}\downarrow} c_{\vec{p}^{\U} \uparrow} {\pm} c_{{-}\vec{p}^{\U}\uparrow} c_{\vec{p} \downarrow} {+} c_{{-}\vec{p}^{\U}\downarrow} c_{\vec{p} \uparrow} \rangle \notag .
\end{align}
Here the spins are either in the singlet or the antiparallel-spin triplet state. 
The gap equation reads
\begin{equation}
 \Delta_{\vec{q} } = - \sum_{\vec{q}'} V_{\tld{\vec{q}}^{\U}, \vec{q}'} \frac{\Delta_{\vec{q}'}}{2\tilde{E}_{\vec{q}'}} \tanh\left(\frac{\tilde{E}_{\vec{q}'}}{2k_B T}\right) \, ,
\end{equation}
$\Delta$ has to satisfy p-wave symmetry, which means 
\begin{equation}
		\Delta_{ (q_y,q_z) } = \Delta^{*}_{ (-q_y,q_z) } = - \Delta^{*}_{ (q_y,-q_z) }.
\label{eq:gapsymmnz}
\end{equation}
This symmetry was used to determine the spin states in Eq.~\eqref{eq:gapnz}. 

We now consider the simplified conditions of Sec.~\ref{sec:quadratic}, where $\vec q^{\U}$ is replaced by $\vec q^{\MU}$, and the electron dispersion is quadratic.
The p-wave solution has the form
\begin{equation}
\delta(x,\varphi) = f(x)\exp(\pm i\varphi) \, .
\end{equation}
Then, we find that $f(x)$ satisfies Eq.~(\ref{eq:gap.quadratic})
with the potential $V(y) \approx -\varepsilon_0/(\sqrt{2\pi}\varepsilon_J) + 1/\sqrt{1 - y^2 }$.
This means that in this simplified model,
the d-wave state considered in the main text and the p-wave solution discussed in this appendix lead approximately to the same critical temperature. 
The p-wave symmetry is energetically slightly preferred, as the constant contribution to the potential $V(y)$ is smaller. 

However, when considering quadratic electron dispersion together with the actual Um\-klapp process,
the non-zero-momentum pairing is strongly suppressed.
This is because the Um\-klapp process does not map states on the Fermi surface to other states on the Fermi surface.
Technically speaking, in Eq.~(\ref{eq:pot_2D}), we would need to replace $(x-x')$ in the expression for the potential $v(x,x',\varphi,\varphi')$ by $(x^U-x')$, where $x^U\varepsilon_0$ is the energy at the Um\-klapp vector $\vec k^U$.
As this will be in most cases far away from the Fermi surface, i.e., $x^U-E_F/\varepsilon_0\gg1$, there will be only small contributions to the integral on the right-hand side of the gap equation \eqref{eq:gap_eeq2D}.

Regarding the tight-binding model, the p-wave gap function must be either discontinuous or zero at the corners of the Fermi surface.
We believe that this suppresses the p-wave solution.

To summarize this appendix, while the simplified model seems to allow for the non-zero-momentum p-wave superconducting pairing, dropping the simplifying assumptions leads to a suppression of this type of pairing both for the quadratic dispersion as well as for the tight-binding model.


\begin{thebibliography}{38}%
\makeatletter
\providecommand \@ifxundefined [1]{%
 \@ifx{#1\undefined}
}%
\providecommand \@ifnum [1]{%
 \ifnum #1\expandafter \@firstoftwo
 \else \expandafter \@secondoftwo
 \fi
}%
\providecommand \@ifx [1]{%
 \ifx #1\expandafter \@firstoftwo
 \else \expandafter \@secondoftwo
 \fi
}%
\providecommand \natexlab [1]{#1}%
\providecommand \enquote  [1]{``#1''}%
\providecommand \bibnamefont  [1]{#1}%
\providecommand \bibfnamefont [1]{#1}%
\providecommand \citenamefont [1]{#1}%
\providecommand \href@noop [0]{\@secondoftwo}%
\providecommand \href [0]{\begingroup \@sanitize@url \@href}%
\providecommand \@href[1]{\@@startlink{#1}\@@href}%
\providecommand \@@href[1]{\endgroup#1\@@endlink}%
\providecommand \@sanitize@url [0]{\catcode `\\12\catcode `\$12\catcode
  `\&12\catcode `\#12\catcode `\^12\catcode `\_12\catcode `\%12\relax}%
\providecommand \@@startlink[1]{}%
\providecommand \@@endlink[0]{}%
\providecommand \url  [0]{\begingroup\@sanitize@url \@url }%
\providecommand \@url [1]{\endgroup\@href {#1}{\urlprefix }}%
\providecommand \urlprefix  [0]{URL }%
\providecommand \Eprint [0]{\href }%
\providecommand \doibase [0]{http://dx.doi.org/}%
\providecommand \selectlanguage [0]{\@gobble}%
\providecommand \bibinfo  [0]{\@secondoftwo}%
\providecommand \bibfield  [0]{\@secondoftwo}%
\providecommand \translation [1]{[#1]}%
\providecommand \BibitemOpen [0]{}%
\providecommand \bibitemStop [0]{}%
\providecommand \bibitemNoStop [0]{.\EOS\space}%
\providecommand \EOS [0]{\spacefactor3000\relax}%
\providecommand \BibitemShut  [1]{\csname bibitem#1\endcsname}%
\let\auto@bib@innerbib\@empty
%</preamble>
\bibitem [{\citenamefont {Jungwirth}\ \emph {et~al.}(2018)\citenamefont
  {Jungwirth}, \citenamefont {Sinova}, \citenamefont {Manchon}, \citenamefont
  {Marti}, \citenamefont {Wunderlich},\ and\ \citenamefont
  {Felser}}]{NatPhys.14.200}%
  \BibitemOpen
  \bibfield  {author} {\bibinfo {author} {\bibfnamefont {T.}~\bibnamefont
  {Jungwirth}}, \bibinfo {author} {\bibfnamefont {J.}~\bibnamefont {Sinova}},
  \bibinfo {author} {\bibfnamefont {A.}~\bibnamefont {Manchon}}, \bibinfo
  {author} {\bibfnamefont {X.}~\bibnamefont {Marti}}, \bibinfo {author}
  {\bibfnamefont {J.}~\bibnamefont {Wunderlich}}, \ and\ \bibinfo {author}
  {\bibfnamefont {C.}~\bibnamefont {Felser}},\ }\href {\doibase
  10.1038/s41567-018-0063-6} {\bibfield  {journal} {\bibinfo  {journal} {Nat.
  Phys.}\ }\textbf {\bibinfo {volume} {14}},\ \bibinfo {pages} {200} (\bibinfo
  {year} {2018})}\BibitemShut {NoStop}%
\bibitem [{\citenamefont {Gomonay}\ \emph {et~al.}(2018)\citenamefont
  {Gomonay}, \citenamefont {Baltz}, \citenamefont {Brataas},\ and\
  \citenamefont {Tserkovnyak}}]{NatPhys.14.213}%
  \BibitemOpen
  \bibfield  {author} {\bibinfo {author} {\bibfnamefont {O.}~\bibnamefont
  {Gomonay}}, \bibinfo {author} {\bibfnamefont {V.}~\bibnamefont {Baltz}},
  \bibinfo {author} {\bibfnamefont {A.}~\bibnamefont {Brataas}}, \ and\
  \bibinfo {author} {\bibfnamefont {Y.}~\bibnamefont {Tserkovnyak}},\ }\href
  {\doibase 10.1038/s41567-018-0063-6} {\bibfield  {journal} {\bibinfo
  {journal} {Nat. Phys.}\ }\textbf {\bibinfo {volume} {14}},\ \bibinfo {pages}
  {213} (\bibinfo {year} {2018})}\BibitemShut {NoStop}%
\bibitem [{\citenamefont {Cheng}\ \emph {et~al.}(2014)\citenamefont {Cheng},
  \citenamefont {Xiao}, \citenamefont {Niu},\ and\ \citenamefont
  {Brataas}}]{PhysRevLett.113.057601}%
  \BibitemOpen
  \bibfield  {author} {\bibinfo {author} {\bibfnamefont {R.}~\bibnamefont
  {Cheng}}, \bibinfo {author} {\bibfnamefont {J.}~\bibnamefont {Xiao}},
  \bibinfo {author} {\bibfnamefont {Q.}~\bibnamefont {Niu}}, \ and\ \bibinfo
  {author} {\bibfnamefont {A.}~\bibnamefont {Brataas}},\ }\href {\doibase
  10.1103/PhysRevLett.113.057601} {\bibfield  {journal} {\bibinfo  {journal}
  {Phys. Rev. Lett.}\ }\textbf {\bibinfo {volume} {113}},\ \bibinfo {pages}
  {057601} (\bibinfo {year} {2014})}\BibitemShut {NoStop}%
\bibitem [{\citenamefont {Takei}\ \emph {et~al.}(2014)\citenamefont {Takei},
  \citenamefont {Halperin}, \citenamefont {Yacoby},\ and\ \citenamefont
  {Tserkovnyak}}]{PhysRevB.90.094408}%
  \BibitemOpen
  \bibfield  {author} {\bibinfo {author} {\bibfnamefont {S.}~\bibnamefont
  {Takei}}, \bibinfo {author} {\bibfnamefont {B.~I.}\ \bibnamefont {Halperin}},
  \bibinfo {author} {\bibfnamefont {A.}~\bibnamefont {Yacoby}}, \ and\ \bibinfo
  {author} {\bibfnamefont {Y.}~\bibnamefont {Tserkovnyak}},\ }\href {\doibase
  10.1103/PhysRevB.90.094408} {\bibfield  {journal} {\bibinfo  {journal} {Phys.
  Rev. B}\ }\textbf {\bibinfo {volume} {90}},\ \bibinfo {pages} {094408}
  (\bibinfo {year} {2014})}\BibitemShut {NoStop}%
\bibitem [{\citenamefont {Fj\ae{}rbu}\ \emph {et~al.}(2017)\citenamefont
  {Fj\ae{}rbu}, \citenamefont {Rohling},\ and\ \citenamefont
  {Brataas}}]{PhysRevB.95.144408}%
  \BibitemOpen
  \bibfield  {author} {\bibinfo {author} {\bibfnamefont {E.~L.}\ \bibnamefont
  {Fj\ae{}rbu}}, \bibinfo {author} {\bibfnamefont {N.}~\bibnamefont {Rohling}},
  \ and\ \bibinfo {author} {\bibfnamefont {A.}~\bibnamefont {Brataas}},\ }\href
  {\doibase 10.1103/PhysRevB.95.144408} {\bibfield  {journal} {\bibinfo
  {journal} {Phys. Rev. B}\ }\textbf {\bibinfo {volume} {95}},\ \bibinfo
  {pages} {144408} (\bibinfo {year} {2017})}\BibitemShut {NoStop}%
\bibitem [{\citenamefont {Cheng}(2014)}]{cheng2014aspects}%
  \BibitemOpen
  \bibfield  {author} {\bibinfo {author} {\bibfnamefont {R.}~\bibnamefont
  {Cheng}},\ }\emph {\bibinfo {title} {Aspects of antiferromagnetic
  spintronics}},\ \href@noop {} {Ph.D. thesis},\ \bibinfo  {school} {The
  University of Texas at Austin} (\bibinfo {year} {2014})\BibitemShut {NoStop}%
\bibitem [{\citenamefont {Moncton}\ \emph {et~al.}(1978)\citenamefont
  {Moncton}, \citenamefont {Shirane}, \citenamefont {Thomlinson}, \citenamefont
  {Ishikawa},\ and\ \citenamefont {Fischer}}]{PhysRevLett.41.1133}%
  \BibitemOpen
  \bibfield  {author} {\bibinfo {author} {\bibfnamefont {D.~E.}\ \bibnamefont
  {Moncton}}, \bibinfo {author} {\bibfnamefont {G.}~\bibnamefont {Shirane}},
  \bibinfo {author} {\bibfnamefont {W.}~\bibnamefont {Thomlinson}}, \bibinfo
  {author} {\bibfnamefont {M.}~\bibnamefont {Ishikawa}}, \ and\ \bibinfo
  {author} {\bibfnamefont {O.}~\bibnamefont {Fischer}},\ }\href {\doibase
  10.1103/PhysRevLett.41.1133} {\bibfield  {journal} {\bibinfo  {journal}
  {Phys. Rev. Lett.}\ }\textbf {\bibinfo {volume} {41}},\ \bibinfo {pages}
  {1133} (\bibinfo {year} {1978})}\BibitemShut {NoStop}%
\bibitem [{\citenamefont {Lynn}\ \emph {et~al.}(2000)\citenamefont {Lynn},
  \citenamefont {Keimer}, \citenamefont {Ulrich}, \citenamefont {Bernhard},\
  and\ \citenamefont {Tallon}}]{PhysRevB.61.R14964}%
  \BibitemOpen
  \bibfield  {author} {\bibinfo {author} {\bibfnamefont {J.~W.}\ \bibnamefont
  {Lynn}}, \bibinfo {author} {\bibfnamefont {B.}~\bibnamefont {Keimer}},
  \bibinfo {author} {\bibfnamefont {C.}~\bibnamefont {Ulrich}}, \bibinfo
  {author} {\bibfnamefont {C.}~\bibnamefont {Bernhard}}, \ and\ \bibinfo
  {author} {\bibfnamefont {J.~L.}\ \bibnamefont {Tallon}},\ }\href {\doibase
  10.1103/PhysRevB.61.R14964} {\bibfield  {journal} {\bibinfo  {journal} {Phys.
  Rev. B}\ }\textbf {\bibinfo {volume} {61}},\ \bibinfo {pages} {R14964}
  (\bibinfo {year} {2000})}\BibitemShut {NoStop}%
\bibitem [{\citenamefont {Aeppli}\ \emph {et~al.}(1988)\citenamefont {Aeppli},
  \citenamefont {Bucher}, \citenamefont {Broholm}, \citenamefont {Kjems},
  \citenamefont {Baumann},\ and\ \citenamefont {Hufnagl}}]{PhysRevLett.60.615}%
  \BibitemOpen
  \bibfield  {author} {\bibinfo {author} {\bibfnamefont {G.}~\bibnamefont
  {Aeppli}}, \bibinfo {author} {\bibfnamefont {E.}~\bibnamefont {Bucher}},
  \bibinfo {author} {\bibfnamefont {C.}~\bibnamefont {Broholm}}, \bibinfo
  {author} {\bibfnamefont {J.~K.}\ \bibnamefont {Kjems}}, \bibinfo {author}
  {\bibfnamefont {J.}~\bibnamefont {Baumann}}, \ and\ \bibinfo {author}
  {\bibfnamefont {J.}~\bibnamefont {Hufnagl}},\ }\href {\doibase
  10.1103/PhysRevLett.60.615} {\bibfield  {journal} {\bibinfo  {journal} {Phys.
  Rev. Lett.}\ }\textbf {\bibinfo {volume} {60}},\ \bibinfo {pages} {615}
  (\bibinfo {year} {1988})}\BibitemShut {NoStop}%
\bibitem [{\citenamefont {Isaacs}\ \emph {et~al.}(1995)\citenamefont {Isaacs},
  \citenamefont {Zschack}, \citenamefont {Broholm}, \citenamefont {Burns},
  \citenamefont {Aeppli}, \citenamefont {Ramirez}, \citenamefont {Palstra},
  \citenamefont {Erwin}, \citenamefont {St\"ucheli},\ and\ \citenamefont
  {Bucher}}]{PhysRevLett.75.1178}%
  \BibitemOpen
  \bibfield  {author} {\bibinfo {author} {\bibfnamefont {E.~D.}\ \bibnamefont
  {Isaacs}}, \bibinfo {author} {\bibfnamefont {P.}~\bibnamefont {Zschack}},
  \bibinfo {author} {\bibfnamefont {C.~L.}\ \bibnamefont {Broholm}}, \bibinfo
  {author} {\bibfnamefont {C.}~\bibnamefont {Burns}}, \bibinfo {author}
  {\bibfnamefont {G.}~\bibnamefont {Aeppli}}, \bibinfo {author} {\bibfnamefont
  {A.~P.}\ \bibnamefont {Ramirez}}, \bibinfo {author} {\bibfnamefont
  {T.~T.~M.}\ \bibnamefont {Palstra}}, \bibinfo {author} {\bibfnamefont
  {R.~W.}\ \bibnamefont {Erwin}}, \bibinfo {author} {\bibfnamefont
  {N.}~\bibnamefont {St\"ucheli}}, \ and\ \bibinfo {author} {\bibfnamefont
  {E.}~\bibnamefont {Bucher}},\ }\href {\doibase 10.1103/PhysRevLett.75.1178}
  {\bibfield  {journal} {\bibinfo  {journal} {Phys. Rev. Lett.}\ }\textbf
  {\bibinfo {volume} {75}},\ \bibinfo {pages} {1178} (\bibinfo {year}
  {1995})}\BibitemShut {NoStop}%
\bibitem [{\citenamefont {Lussier}\ \emph {et~al.}(1997)\citenamefont
  {Lussier}, \citenamefont {Mao}, \citenamefont {Schr\"oder}, \citenamefont
  {Garrett}, \citenamefont {Gaulin}, \citenamefont {Shapiro},\ and\
  \citenamefont {Buyers}}]{PhysRevB.56.11749}%
  \BibitemOpen
  \bibfield  {author} {\bibinfo {author} {\bibfnamefont {J.~G.}\ \bibnamefont
  {Lussier}}, \bibinfo {author} {\bibfnamefont {M.}~\bibnamefont {Mao}},
  \bibinfo {author} {\bibfnamefont {A.}~\bibnamefont {Schr\"oder}}, \bibinfo
  {author} {\bibfnamefont {J.~D.}\ \bibnamefont {Garrett}}, \bibinfo {author}
  {\bibfnamefont {B.~D.}\ \bibnamefont {Gaulin}}, \bibinfo {author}
  {\bibfnamefont {S.~M.}\ \bibnamefont {Shapiro}}, \ and\ \bibinfo {author}
  {\bibfnamefont {W.~J.~L.}\ \bibnamefont {Buyers}},\ }\href {\doibase
  10.1103/PhysRevB.56.11749} {\bibfield  {journal} {\bibinfo  {journal} {Phys.
  Rev. B}\ }\textbf {\bibinfo {volume} {56}},\ \bibinfo {pages} {11749}
  (\bibinfo {year} {1997})}\BibitemShut {NoStop}%
\bibitem [{\citenamefont {Lee}\ \emph {et~al.}(2006)\citenamefont {Lee},
  \citenamefont {Nagaosa},\ and\ \citenamefont {Wen}}]{RevModPhys.78.17}%
  \BibitemOpen
  \bibfield  {author} {\bibinfo {author} {\bibfnamefont {P.~A.}\ \bibnamefont
  {Lee}}, \bibinfo {author} {\bibfnamefont {N.}~\bibnamefont {Nagaosa}}, \ and\
  \bibinfo {author} {\bibfnamefont {X.-G.}\ \bibnamefont {Wen}},\ }\href
  {\doibase 10.1103/RevModPhys.78.17} {\bibfield  {journal} {\bibinfo
  {journal} {Rev. Mod. Phys.}\ }\textbf {\bibinfo {volume} {78}},\ \bibinfo
  {pages} {17} (\bibinfo {year} {2006})}\BibitemShut {NoStop}%
\bibitem [{\citenamefont {Ni}\ \emph {et~al.}(2008)\citenamefont {Ni},
  \citenamefont {Tillman}, \citenamefont {Yan}, \citenamefont {Kracher},
  \citenamefont {Hannahs}, \citenamefont {Bud'ko},\ and\ \citenamefont
  {Canfield}}]{PhysRevB.78.214515}%
  \BibitemOpen
  \bibfield  {author} {\bibinfo {author} {\bibfnamefont {N.}~\bibnamefont
  {Ni}}, \bibinfo {author} {\bibfnamefont {M.~E.}\ \bibnamefont {Tillman}},
  \bibinfo {author} {\bibfnamefont {J.-Q.}\ \bibnamefont {Yan}}, \bibinfo
  {author} {\bibfnamefont {A.}~\bibnamefont {Kracher}}, \bibinfo {author}
  {\bibfnamefont {S.~T.}\ \bibnamefont {Hannahs}}, \bibinfo {author}
  {\bibfnamefont {S.~L.}\ \bibnamefont {Bud'ko}}, \ and\ \bibinfo {author}
  {\bibfnamefont {P.~C.}\ \bibnamefont {Canfield}},\ }\href {\doibase
  10.1103/PhysRevB.78.214515} {\bibfield  {journal} {\bibinfo  {journal} {Phys.
  Rev. B}\ }\textbf {\bibinfo {volume} {78}},\ \bibinfo {pages} {214515}
  (\bibinfo {year} {2008})}\BibitemShut {NoStop}%
\bibitem [{\citenamefont {Chu}\ \emph {et~al.}(2009)\citenamefont {Chu},
  \citenamefont {Analytis}, \citenamefont {Kucharczyk},\ and\ \citenamefont
  {Fisher}}]{PhysRevB.79.014506}%
  \BibitemOpen
  \bibfield  {author} {\bibinfo {author} {\bibfnamefont {J.-H.}\ \bibnamefont
  {Chu}}, \bibinfo {author} {\bibfnamefont {J.~G.}\ \bibnamefont {Analytis}},
  \bibinfo {author} {\bibfnamefont {C.}~\bibnamefont {Kucharczyk}}, \ and\
  \bibinfo {author} {\bibfnamefont {I.~R.}\ \bibnamefont {Fisher}},\ }\href
  {\doibase 10.1103/PhysRevB.79.014506} {\bibfield  {journal} {\bibinfo
  {journal} {Phys. Rev. B}\ }\textbf {\bibinfo {volume} {79}},\ \bibinfo
  {pages} {014506} (\bibinfo {year} {2009})}\BibitemShut {NoStop}%
\bibitem [{\citenamefont {Rotter}\ \emph {et~al.}(2009)\citenamefont {Rotter},
  \citenamefont {Tegel}, \citenamefont {Schellenberg}, \citenamefont
  {Schappacher}, \citenamefont {P{\"o}ttgen}, \citenamefont {Deisenhofer},
  \citenamefont {G{\"u}nther}, \citenamefont {Schrettle}, \citenamefont
  {Loidl},\ and\ \citenamefont {Johrendt}}]{NewJPhys2009Rotter}%
  \BibitemOpen
  \bibfield  {author} {\bibinfo {author} {\bibfnamefont {M.}~\bibnamefont
  {Rotter}}, \bibinfo {author} {\bibfnamefont {M.}~\bibnamefont {Tegel}},
  \bibinfo {author} {\bibfnamefont {I.}~\bibnamefont {Schellenberg}}, \bibinfo
  {author} {\bibfnamefont {F.~M.}\ \bibnamefont {Schappacher}}, \bibinfo
  {author} {\bibfnamefont {R.}~\bibnamefont {P{\"o}ttgen}}, \bibinfo {author}
  {\bibfnamefont {J.}~\bibnamefont {Deisenhofer}}, \bibinfo {author}
  {\bibfnamefont {A.}~\bibnamefont {G{\"u}nther}}, \bibinfo {author}
  {\bibfnamefont {F.}~\bibnamefont {Schrettle}}, \bibinfo {author}
  {\bibfnamefont {A.}~\bibnamefont {Loidl}}, \ and\ \bibinfo {author}
  {\bibfnamefont {D.}~\bibnamefont {Johrendt}},\ }\href
  {http://stacks.iop.org/1367-2630/11/i=2/a=025014} {\bibfield  {journal}
  {\bibinfo  {journal} {New J. Phys.}\ }\textbf {\bibinfo {volume} {11}},\
  \bibinfo {pages} {025014} (\bibinfo {year} {2009})}\BibitemShut {NoStop}%
\bibitem [{\citenamefont {Liu}\ \emph {et~al.}(2008)\citenamefont {Liu},
  \citenamefont {Wu}, \citenamefont {Wu}, \citenamefont {Fang}, \citenamefont
  {Chen}, \citenamefont {Li}, \citenamefont {Liu}, \citenamefont {Xie},
  \citenamefont {Wang}, \citenamefont {Yang}, \citenamefont {Ding},
  \citenamefont {He}, \citenamefont {Feng},\ and\ \citenamefont
  {Chen}}]{PhysRevLett.101.087001}%
  \BibitemOpen
  \bibfield  {author} {\bibinfo {author} {\bibfnamefont {R.~H.}\ \bibnamefont
  {Liu}}, \bibinfo {author} {\bibfnamefont {G.}~\bibnamefont {Wu}}, \bibinfo
  {author} {\bibfnamefont {T.}~\bibnamefont {Wu}}, \bibinfo {author}
  {\bibfnamefont {D.~F.}\ \bibnamefont {Fang}}, \bibinfo {author}
  {\bibfnamefont {H.}~\bibnamefont {Chen}}, \bibinfo {author} {\bibfnamefont
  {S.~Y.}\ \bibnamefont {Li}}, \bibinfo {author} {\bibfnamefont
  {K.}~\bibnamefont {Liu}}, \bibinfo {author} {\bibfnamefont {Y.~L.}\
  \bibnamefont {Xie}}, \bibinfo {author} {\bibfnamefont {X.~F.}\ \bibnamefont
  {Wang}}, \bibinfo {author} {\bibfnamefont {R.~L.}\ \bibnamefont {Yang}},
  \bibinfo {author} {\bibfnamefont {L.}~\bibnamefont {Ding}}, \bibinfo {author}
  {\bibfnamefont {C.}~\bibnamefont {He}}, \bibinfo {author} {\bibfnamefont
  {D.~L.}\ \bibnamefont {Feng}}, \ and\ \bibinfo {author} {\bibfnamefont
  {X.~H.}\ \bibnamefont {Chen}},\ }\href {\doibase
  10.1103/PhysRevLett.101.087001} {\bibfield  {journal} {\bibinfo  {journal}
  {Phys. Rev. Lett.}\ }\textbf {\bibinfo {volume} {101}},\ \bibinfo {pages}
  {087001} (\bibinfo {year} {2008})}\BibitemShut {NoStop}%
\bibitem [{\citenamefont {Karchev}(2017)}]{PhysRevB.96.214409}%
  \BibitemOpen
  \bibfield  {author} {\bibinfo {author} {\bibfnamefont {N.}~\bibnamefont
  {Karchev}},\ }\href {\doibase 10.1103/PhysRevB.96.214409} {\bibfield
  {journal} {\bibinfo  {journal} {Phys. Rev. B}\ }\textbf {\bibinfo {volume}
  {96}},\ \bibinfo {pages} {214409} (\bibinfo {year} {2017})}\BibitemShut
  {NoStop}%
\bibitem [{\citenamefont {Shimahara}(1994)}]{JPSJ.63.1861}%
  \BibitemOpen
  \bibfield  {author} {\bibinfo {author} {\bibfnamefont {H.}~\bibnamefont
  {Shimahara}},\ }\href {\doibase 10.1143/JPSJ.63.1861} {\bibfield  {journal}
  {\bibinfo  {journal} {J. Phys. Soc. Jpn.}\ }\textbf {\bibinfo {volume}
  {63}},\ \bibinfo {pages} {1861} (\bibinfo {year} {1994})}\BibitemShut
  {NoStop}%
\bibitem [{\citenamefont {kar}\ \emph {et~al.}(2018)\citenamefont {kar},
  \citenamefont {Paul},\ and\ \citenamefont {Misra}}]{KAR201818}%
  \BibitemOpen
  \bibfield  {author} {\bibinfo {author} {\bibfnamefont {R.}~\bibnamefont
  {kar}}, \bibinfo {author} {\bibfnamefont {B.~C.}\ \bibnamefont {Paul}}, \
  and\ \bibinfo {author} {\bibfnamefont {A.}~\bibnamefont {Misra}},\
  }\href@noop {} {\bibfield  {journal} {\bibinfo  {journal} {Physica C
  (Amsterdam)}\ }\textbf {\bibinfo {volume} {545}},\ \bibinfo {pages} {18}
  (\bibinfo {year} {2018})}\BibitemShut {NoStop}%
\bibitem [{\citenamefont {Wu}\ and\ \citenamefont
  {Phillips}(2011)}]{WuJPC2011}%
  \BibitemOpen
  \bibfield  {author} {\bibinfo {author} {\bibfnamefont {J.}~\bibnamefont
  {Wu}}\ and\ \bibinfo {author} {\bibfnamefont {P.}~\bibnamefont {Phillips}},\
  }\href@noop {} {\bibfield  {journal} {\bibinfo  {journal} {J. Phys.: Condens.
  Matter}\ }\textbf {\bibinfo {volume} {23}},\ \bibinfo {pages} {094203}
  (\bibinfo {year} {2011})}\BibitemShut {NoStop}%
\bibitem [{\citenamefont {Kargarian}\ \emph {et~al.}(2016)\citenamefont
  {Kargarian}, \citenamefont {Efimkin},\ and\ \citenamefont
  {Galitski}}]{KagarianPRL2016}%
  \BibitemOpen
  \bibfield  {author} {\bibinfo {author} {\bibfnamefont {M.}~\bibnamefont
  {Kargarian}}, \bibinfo {author} {\bibfnamefont {D.~K.}\ \bibnamefont
  {Efimkin}}, \ and\ \bibinfo {author} {\bibfnamefont {V.}~\bibnamefont
  {Galitski}},\ }\href {\doibase 10.1103/PhysRevLett.117.076806} {\bibfield
  {journal} {\bibinfo  {journal} {Phys. Rev. Lett.}\ }\textbf {\bibinfo
  {volume} {117}},\ \bibinfo {pages} {076806} (\bibinfo {year}
  {2016})}\BibitemShut {NoStop}%
\bibitem [{\citenamefont {Gong}\ \emph {et~al.}(2017)\citenamefont {Gong},
  \citenamefont {Kargarian}, \citenamefont {Stern}, \citenamefont {Yue},
  \citenamefont {Zhou}, \citenamefont {Jin}, \citenamefont {Galitski},
  \citenamefont {Yakovenko},\ and\ \citenamefont {Xia}}]{GongeSCIENCE2017}%
  \BibitemOpen
  \bibfield  {author} {\bibinfo {author} {\bibfnamefont {X.}~\bibnamefont
  {Gong}}, \bibinfo {author} {\bibfnamefont {M.}~\bibnamefont {Kargarian}},
  \bibinfo {author} {\bibfnamefont {A.}~\bibnamefont {Stern}}, \bibinfo
  {author} {\bibfnamefont {D.}~\bibnamefont {Yue}}, \bibinfo {author}
  {\bibfnamefont {H.}~\bibnamefont {Zhou}}, \bibinfo {author} {\bibfnamefont
  {X.}~\bibnamefont {Jin}}, \bibinfo {author} {\bibfnamefont {V.~M.}\
  \bibnamefont {Galitski}}, \bibinfo {author} {\bibfnamefont {V.~M.}\
  \bibnamefont {Yakovenko}}, \ and\ \bibinfo {author} {\bibfnamefont
  {J.}~\bibnamefont {Xia}},\ }\href {\doibase 10.1126/sciadv.1602579}
  {\bibfield  {journal} {\bibinfo  {journal} {Science Advances}\ }\textbf
  {\bibinfo {volume} {3}} (\bibinfo {year} {2017}),\
  10.1126/sciadv.1602579}\BibitemShut {NoStop}%
\bibitem [{\citenamefont {Hugdal}\ \emph {et~al.}(2018)\citenamefont {Hugdal},
  \citenamefont {Rex}, \citenamefont {Nogueira},\ and\ \citenamefont
  {Sudb\o{}}}]{hugdal2018magnon}%
  \BibitemOpen
  \bibfield  {author} {\bibinfo {author} {\bibfnamefont {H.~G.}\ \bibnamefont
  {Hugdal}}, \bibinfo {author} {\bibfnamefont {S.}~\bibnamefont {Rex}},
  \bibinfo {author} {\bibfnamefont {F.~S.}\ \bibnamefont {Nogueira}}, \ and\
  \bibinfo {author} {\bibfnamefont {A.}~\bibnamefont {Sudb\o{}}},\ }\href
  {\doibase 10.1103/PhysRevB.97.195438} {\bibfield  {journal} {\bibinfo
  {journal} {Phys. Rev. B}\ }\textbf {\bibinfo {volume} {97}},\ \bibinfo
  {pages} {195438} (\bibinfo {year} {2018})}\BibitemShut {NoStop}%
\bibitem [{\citenamefont {Rohling}\ \emph {et~al.}(2018)\citenamefont
  {Rohling}, \citenamefont {Fj\ae{}rbu},\ and\ \citenamefont
  {Brataas}}]{PhysRevB.97.115401}%
  \BibitemOpen
  \bibfield  {author} {\bibinfo {author} {\bibfnamefont {N.}~\bibnamefont
  {Rohling}}, \bibinfo {author} {\bibfnamefont {E.~L.}\ \bibnamefont
  {Fj\ae{}rbu}}, \ and\ \bibinfo {author} {\bibfnamefont {A.}~\bibnamefont
  {Brataas}},\ }\href {\doibase 10.1103/PhysRevB.97.115401} {\bibfield
  {journal} {\bibinfo  {journal} {Phys. Rev. B}\ }\textbf {\bibinfo {volume}
  {97}},\ \bibinfo {pages} {115401} (\bibinfo {year} {2018})}\BibitemShut
  {NoStop}%
\bibitem [{\citenamefont {Seki}\ \emph {et~al.}(2015)\citenamefont {Seki},
  \citenamefont {Ideue}, \citenamefont {Kubota}, \citenamefont {Kozuka},
  \citenamefont {Takagi}, \citenamefont {Nakamura}, \citenamefont {Kaneko},
  \citenamefont {Kawasaki},\ and\ \citenamefont
  {Tokura}}]{PhysRevLett.115.266601}%
  \BibitemOpen
  \bibfield  {author} {\bibinfo {author} {\bibfnamefont {S.}~\bibnamefont
  {Seki}}, \bibinfo {author} {\bibfnamefont {T.}~\bibnamefont {Ideue}},
  \bibinfo {author} {\bibfnamefont {M.}~\bibnamefont {Kubota}}, \bibinfo
  {author} {\bibfnamefont {Y.}~\bibnamefont {Kozuka}}, \bibinfo {author}
  {\bibfnamefont {R.}~\bibnamefont {Takagi}}, \bibinfo {author} {\bibfnamefont
  {M.}~\bibnamefont {Nakamura}}, \bibinfo {author} {\bibfnamefont
  {Y.}~\bibnamefont {Kaneko}}, \bibinfo {author} {\bibfnamefont
  {M.}~\bibnamefont {Kawasaki}}, \ and\ \bibinfo {author} {\bibfnamefont
  {Y.}~\bibnamefont {Tokura}},\ }\href {\doibase
  10.1103/PhysRevLett.115.266601} {\bibfield  {journal} {\bibinfo  {journal}
  {Phys. Rev. Lett.}\ }\textbf {\bibinfo {volume} {115}},\ \bibinfo {pages}
  {266601} (\bibinfo {year} {2015})}\BibitemShut {NoStop}%
\bibitem [{\citenamefont {Wu}\ \emph {et~al.}(2016)\citenamefont {Wu},
  \citenamefont {Zhang}, \citenamefont {KC}, \citenamefont {Borisov},
  \citenamefont {Pearson}, \citenamefont {Jiang}, \citenamefont {Lederman},
  \citenamefont {Hoffmann},\ and\ \citenamefont
  {Bhattacharya}}]{PhysRevLett.116.097204}%
  \BibitemOpen
  \bibfield  {author} {\bibinfo {author} {\bibfnamefont {S.~M.}\ \bibnamefont
  {Wu}}, \bibinfo {author} {\bibfnamefont {W.}~\bibnamefont {Zhang}}, \bibinfo
  {author} {\bibfnamefont {A.}~\bibnamefont {KC}}, \bibinfo {author}
  {\bibfnamefont {P.}~\bibnamefont {Borisov}}, \bibinfo {author} {\bibfnamefont
  {J.~E.}\ \bibnamefont {Pearson}}, \bibinfo {author} {\bibfnamefont {J.~S.}\
  \bibnamefont {Jiang}}, \bibinfo {author} {\bibfnamefont {D.}~\bibnamefont
  {Lederman}}, \bibinfo {author} {\bibfnamefont {A.}~\bibnamefont {Hoffmann}},
  \ and\ \bibinfo {author} {\bibfnamefont {A.}~\bibnamefont {Bhattacharya}},\
  }\href {\doibase 10.1103/PhysRevLett.116.097204} {\bibfield  {journal}
  {\bibinfo  {journal} {Phys. Rev. Lett.}\ }\textbf {\bibinfo {volume} {116}},\
  \bibinfo {pages} {097204} (\bibinfo {year} {2016})}\BibitemShut {NoStop}%
\bibitem [{\citenamefont {Ross}\ \emph {et~al.}(2015)\citenamefont {Ross},
  \citenamefont {Schreier}, \citenamefont {Lotze}, \citenamefont {Huebl},
  \citenamefont {Gross},\ and\ \citenamefont
  {Goennenwein}}]{JAppPhys.118.233907}%
  \BibitemOpen
  \bibfield  {author} {\bibinfo {author} {\bibfnamefont {P.}~\bibnamefont
  {Ross}}, \bibinfo {author} {\bibfnamefont {M.}~\bibnamefont {Schreier}},
  \bibinfo {author} {\bibfnamefont {J.}~\bibnamefont {Lotze}}, \bibinfo
  {author} {\bibfnamefont {H.}~\bibnamefont {Huebl}}, \bibinfo {author}
  {\bibfnamefont {R.}~\bibnamefont {Gross}}, \ and\ \bibinfo {author}
  {\bibfnamefont {S.~T.~B.}\ \bibnamefont {Goennenwein}},\ }\href {\doibase
  10.1063/1.4937913} {\bibfield  {journal} {\bibinfo  {journal} {J. App.
  Phys.}\ }\textbf {\bibinfo {volume} {118}},\ \bibinfo {pages} {233907}
  (\bibinfo {year} {2015})}\BibitemShut {NoStop}%
\bibitem [{\citenamefont {Wang}\ \emph {et~al.}(2014)\citenamefont {Wang},
  \citenamefont {Du}, \citenamefont {Hammel},\ and\ \citenamefont
  {Yang}}]{PhysRevLett.113.097202}%
  \BibitemOpen
  \bibfield  {author} {\bibinfo {author} {\bibfnamefont {H.}~\bibnamefont
  {Wang}}, \bibinfo {author} {\bibfnamefont {C.}~\bibnamefont {Du}}, \bibinfo
  {author} {\bibfnamefont {P.~C.}\ \bibnamefont {Hammel}}, \ and\ \bibinfo
  {author} {\bibfnamefont {F.}~\bibnamefont {Yang}},\ }\href {\doibase
  10.1103/PhysRevLett.113.097202} {\bibfield  {journal} {\bibinfo  {journal}
  {Phys. Rev. Lett.}\ }\textbf {\bibinfo {volume} {113}},\ \bibinfo {pages}
  {097202} (\bibinfo {year} {2014})}\BibitemShut {NoStop}%
\bibitem [{\citenamefont {Erlandsen}\ \emph {et~al.}(2019)\citenamefont
  {Erlandsen}, \citenamefont {Kamra}, \citenamefont {Brataas},\ and\
  \citenamefont {Sudb\o{}}}]{Erlandsen2019}%
  \BibitemOpen
  \bibfield  {author} {\bibinfo {author} {\bibfnamefont {E.}~\bibnamefont
  {Erlandsen}}, \bibinfo {author} {\bibfnamefont {A.}~\bibnamefont {Kamra}},
  \bibinfo {author} {\bibfnamefont {A.}~\bibnamefont {Brataas}}, \ and\
  \bibinfo {author} {\bibfnamefont {A.}~\bibnamefont {Sudb\o{}}},\ }\href
  {https://arxiv.org/abs/1903.01470} {\bibfield  {journal} {\bibinfo  {journal}
  {arXiv:1903.01470}\ } (\bibinfo {year} {2019})}\BibitemShut {NoStop}%
\bibitem [{\citenamefont {Jauch}\ \emph {et~al.}(1983)\citenamefont {Jauch},
  \citenamefont {Schneider},\ and\ \citenamefont {Dachs}}]{JAUCH1983907}%
  \BibitemOpen
  \bibfield  {author} {\bibinfo {author} {\bibfnamefont {W.}~\bibnamefont
  {Jauch}}, \bibinfo {author} {\bibfnamefont {J.}~\bibnamefont {Schneider}}, \
  and\ \bibinfo {author} {\bibfnamefont {H.}~\bibnamefont {Dachs}},\ }\href
  {\doibase https://doi.org/10.1016/0038-1098(83)90146-1} {\bibfield  {journal}
  {\bibinfo  {journal} {Solid State Commun.}\ }\textbf {\bibinfo {volume}
  {48}},\ \bibinfo {pages} {907 } (\bibinfo {year} {1983})}\BibitemShut
  {NoStop}%
\bibitem [{\citenamefont {Low}\ \emph {et~al.}(1964)\citenamefont {Low},
  \citenamefont {Okazaki}, \citenamefont {Stevenson},\ and\ \citenamefont
  {Turberfield}}]{JAppPhys.35.998}%
  \BibitemOpen
  \bibfield  {author} {\bibinfo {author} {\bibfnamefont {G.~G.}\ \bibnamefont
  {Low}}, \bibinfo {author} {\bibfnamefont {A.}~\bibnamefont {Okazaki}},
  \bibinfo {author} {\bibfnamefont {R.~W.~H.}\ \bibnamefont {Stevenson}}, \
  and\ \bibinfo {author} {\bibfnamefont {K.~C.}\ \bibnamefont {Turberfield}},\
  }\href {\doibase 10.1063/1.1713575} {\bibfield  {journal} {\bibinfo
  {journal} {J. App. Phys.}\ }\textbf {\bibinfo {volume} {35}},\ \bibinfo
  {pages} {998} (\bibinfo {year} {1964})}\BibitemShut {NoStop}%
\bibitem [{\citenamefont {Tserkovnyak}\ \emph {et~al.}(2005)\citenamefont
  {Tserkovnyak}, \citenamefont {Brataas}, \citenamefont {Bauer},\ and\
  \citenamefont {Halperin}}]{TserkovnyakRMP2005}%
  \BibitemOpen
  \bibfield  {author} {\bibinfo {author} {\bibfnamefont {Y.}~\bibnamefont
  {Tserkovnyak}}, \bibinfo {author} {\bibfnamefont {A.}~\bibnamefont
  {Brataas}}, \bibinfo {author} {\bibfnamefont {G.~E.~W.}\ \bibnamefont
  {Bauer}}, \ and\ \bibinfo {author} {\bibfnamefont {B.~I.}\ \bibnamefont
  {Halperin}},\ }\href {https://doi.org/10.1103/RevModPhys.77.1375} {\bibfield
  {journal} {\bibinfo  {journal} {Rev. Mod. Phys.}\ }\textbf {\bibinfo {volume}
  {77}},\ \bibinfo {pages} {1375} (\bibinfo {year} {2005})}\BibitemShut
  {NoStop}%
\bibitem [{\citenamefont {Ashcroft}\ and\ \citenamefont
  {Mermin}(1976)}]{ashcroft1976}%
  \BibitemOpen
  \bibfield  {author} {\bibinfo {author} {\bibfnamefont {N.~W.}\ \bibnamefont
  {Ashcroft}}\ and\ \bibinfo {author} {\bibfnamefont {N.~D.}\ \bibnamefont
  {Mermin}},\ }\href {http://cds.cern.ch/record/102652} {\emph {\bibinfo
  {title} {Solid state physics}}}\ (\bibinfo  {publisher} {Holt,Rinehart and
  Winston, New York},\ \bibinfo {year} {1976})\BibitemShut {NoStop}%
\bibitem [{Note1()}]{Note1}%
  \BibitemOpen
  \bibinfo {note} {The exchange coupling as given in Ref.~\cite
  {Miao_et_al_NatComm2013} corresponds to $J_I/2$.}\BibitemShut {Stop}%
\bibitem [{\citenamefont {Tkaczyk}(1988)}]{Tkaczyk_thesis}%
  \BibitemOpen
  \bibfield  {author} {\bibinfo {author} {\bibfnamefont {J.~E.}\ \bibnamefont
  {Tkaczyk}},\ }\href {http://hdl.handle.net/1721.1/17231} {Ph.D. thesis},\
  \bibinfo  {school} {MIT} (\bibinfo {year} {1988})\BibitemShut {NoStop}%
\bibitem [{\citenamefont {Roesler}\ \emph {et~al.}(1994)\citenamefont
  {Roesler}, \citenamefont {Filipkowski}, \citenamefont {Broussard},
  \citenamefont {Idzerda}, \citenamefont {Osofsky},\ and\ \citenamefont
  {Soulen}}]{Roesler_et_al_1994}%
  \BibitemOpen
  \bibfield  {author} {\bibinfo {author} {\bibfnamefont {G.~M.}\ \bibnamefont
  {Roesler}}, \bibinfo {author} {\bibfnamefont {M.~E.}\ \bibnamefont
  {Filipkowski}}, \bibinfo {author} {\bibfnamefont {P.~R.}\ \bibnamefont
  {Broussard}}, \bibinfo {author} {\bibfnamefont {Y.~U.}\ \bibnamefont
  {Idzerda}}, \bibinfo {author} {\bibfnamefont {M.~S.}\ \bibnamefont
  {Osofsky}}, \ and\ \bibinfo {author} {\bibfnamefont {R.~J.}\ \bibnamefont
  {Soulen}},\ }\href {http://dx.doi.org/10.1117/12.179175} {\bibfield
  {journal} {\bibinfo  {journal} {Proc.~SPIE}\ }\textbf {\bibinfo {volume}
  {2157}},\ \bibinfo {pages} {285} (\bibinfo {year} {1994})}\BibitemShut
  {NoStop}%
\bibitem [{\citenamefont {Miao}\ \emph {et~al.}(2014)\citenamefont {Miao},
  \citenamefont {Chang}, \citenamefont {Assaf}, \citenamefont {Heiman},\ and\
  \citenamefont {Moodera}}]{Miao_et_al_NatComm2013}%
  \BibitemOpen
  \bibfield  {author} {\bibinfo {author} {\bibfnamefont {G.-X.}\ \bibnamefont
  {Miao}}, \bibinfo {author} {\bibfnamefont {J.}~\bibnamefont {Chang}},
  \bibinfo {author} {\bibfnamefont {B.~A.}\ \bibnamefont {Assaf}}, \bibinfo
  {author} {\bibfnamefont {D.}~\bibnamefont {Heiman}}, \ and\ \bibinfo {author}
  {\bibfnamefont {J.~S.}\ \bibnamefont {Moodera}},\ }\href
  {http://dx.doi.org/10.1038/ncomms4682} {\bibfield  {journal} {\bibinfo
  {journal} {Nature Communications}\ }\textbf {\bibinfo {volume} {5}},\
  \bibinfo {pages} {3682} (\bibinfo {year} {2014})}\BibitemShut {NoStop}%
\bibitem [{\citenamefont {Nass}\ \emph {et~al.}(1981)\citenamefont {Nass},
  \citenamefont {Levin},\ and\ \citenamefont {Grest}}]{PhysRevLett.46.614}%
  \BibitemOpen
  \bibfield  {author} {\bibinfo {author} {\bibfnamefont {M.~J.}\ \bibnamefont
  {Nass}}, \bibinfo {author} {\bibfnamefont {K.}~\bibnamefont {Levin}}, \ and\
  \bibinfo {author} {\bibfnamefont {G.~S.}\ \bibnamefont {Grest}},\ }\href
  {\doibase 10.1103/PhysRevLett.46.614} {\bibfield  {journal} {\bibinfo
  {journal} {Phys. Rev. Lett.}\ }\textbf {\bibinfo {volume} {46}},\ \bibinfo
  {pages} {614} (\bibinfo {year} {1981})}\BibitemShut {NoStop}%
\end{thebibliography}
\end{document}